\documentclass[a4paper,11pt]{article}
\pdfoutput=1 
\usepackage{jcappub} 
\usepackage[T1]{fontenc}

\usepackage{amsmath, amssymb}
\usepackage{bm}
\usepackage{graphicx,color}
\usepackage{soul}
\usepackage{layouts}
\usepackage{multirow}
\usepackage{makecell}
\usepackage{mathtools}
\usepackage{bbold}
\usepackage{hyperref}

\usepackage{tikz}
\usetikzlibrary{arrows.meta}

\renewcommand{\vec}[1]{\boldsymbol{\bm{#1}}}
\newcommand{\vecs}[1]{\tilde{\vec{#1}}}
\newcommand{\tracer}{\mathrm{g}}
\newcommand{\dgd}{\delta_{\tracer,\mathrm{det}}}
\newcommand{\dgds}{\tilde{\delta}_{\tracer,\mathrm{det}}}
\newcommand{\udgpp}{u_{\tracer,\mathrm{det}\pp}}
\newcommand{\delaunay}{\mathrm{G,D}}
\newcommand{\pp}{\parallel}
\newcommand{\op}{\mathcal{O}}
\newcommand{\uop}{\mathcal{U}}
\newcommand{\los}{\hat{\vec{n}}}
\newcommand{\dirac}{\mathrm{D}}
\newcommand{\lagrangian}{\mathrm{L}}

\newcommand{\nyquist}{\mathrm{Ny}}
\newcommand{\grid}{\mathrm{G}}
\newcommand{\lin}{^{(1)}}
\newcommand{\lpt}{\mathrm{LPT}}
\newcommand{\ini}{\mathrm{G,in}}
\newcommand{\fwd}{\mathrm{G,fwd}}
\newcommand{\eul}{\mathrm{G,Eul}}

\newcommand{\lh}{\mathrm{G,LH}}
\newcommand{\mpc}{\mathrm{Mpc}}
\newcommand{\tr}{\mathrm{tr}}

\newcommand{\LO}{\mathrm{L.O.}}
\newcommand{\bias}{\mathrm{bias}}
\newcommand{\leftfield}{\texttt{LEFTfield}}
\newcommand{\matter}{\mathrm{m}}

\def\refeq#1{eq.~\eqref{eq:#1}}

\title{Fast, Accurate and Perturbative Forward Modeling of Galaxy Clustering \\[.5\baselineskip] Part II: Redshift Space}

\author[a,b]{Julia Stadler,}
\author[a,b]{Fabian Schmidt,}
\author[a]{Martin Reinecke} 
\author[c]{and Matteo Esposito}
\affiliation[a]{Max-Planck-Institut für Astrophysik, Karl-Schwarzschild-Str. 1, 85748 Garching, Germany}
\affiliation[b]{Excellence Cluster ORIGINS, Boltzmannstr. 2, 85748 Garching, Germany}
\affiliation[c]{Max Planck Institute for extraterrestrial Physics, Giessenbachstraße 1, 85748 Garching, Germany}
\emailAdd{jstadler@mpa-garching.mpg.de}
\emailAdd{fabians@mpa-garching.mpg.de}
\emailAdd{martin@mpa-garching.mpg.de}
\emailAdd{esposito@mpe.mpg.de}

\abstract{
Forward modeling the galaxy density within the Effective Field Theory of Large Scale Structure (EFT of LSS) enables field-level analyses that are robust to theoretical uncertainties. At the same time, they can maximize the constraining power from galaxy clustering on the scales amenable to perturbation theory. In order to apply the method to galaxy surveys, the forward model must account for the full observational complexity of the data. In this context, a major challenge is the inclusion of redshift space distortions (RSDs) from the peculiar motion of galaxies. Here, we present improvements in the efficiency and accuracy of the RSD modeling in the perturbative \leftfield\ forward model. We perform a detailed quantification of the perturbative and numerical error for the prediction of momentum, velocity and the redshift-space matter density. Further, we test the recovery of cosmological parameters at the field level, namely the growth rate $f$, from simulated halos in redshift space. For a rigorous test and to scan through a wide range of analysis choices, we  fix the linear (initial) density field to the known ground truth but marginalize over all unknown bias coefficients and noise amplitudes. With a third-order model for gravity and bias, our results yield $<1\,\%$ statistical and $<1.5\,\%$ systematic error. The computational cost of the redshift-space forward model is only $\sim 1.5$ times of the rest frame equivalent, enabling future field-level inference that simultaneously targets cosmological parameters and the initial matter distribution. 
}

\begin{document}
\maketitle
\flushbottom

\section{Introduction}
\label{sec:intro}

Perturbative forward modeling is a powerful tool to analyze galaxy clustering, in particular in light of a new generation of spectroscopic galaxy surveys such as DESI \cite{DESI:2016fyo}, Euclid \cite{Amendola:2016saw, Euclid:2024yrr}, PFS \cite{2014PASJ...66R...1T} or SPHEREx \cite{Dore:2014cca}. The measurement of the galaxy density at unprecedented volume and depth implies that more information can be gained from non-Gaussian features on quasi-linear scales, but it also demands very accurate theoretical models. In this context, the Effective Field Theory of Large Scale Structure (EFT of LSS, \cite{Baumann:2010tm, Carrasco:2012cv}) provides rigorous control over theoretical uncertainties in the galaxy bias relation \cite{Desjacques:2016bnm}, and a forward model of the three-dimensional galaxy density can access non-Gaussian information either through \emph{higher-order summaries}, via simulation-based inference (SBI, \cite{2020PNAS..11730055C}) with \cite{Hahn:2023udg, Tucci:2023bag}, or by a full \emph{field-level analysis} \cite{Schmidt:2018bkr, Jasche:2018oym, Lavaux:2019fjr, 2013MNRAS.432..894J, 2013MNRAS.429L..84K, 2013ApJ...772...63W, Wang:2014hia, Modi:2018cfi, Shallue:2022mhf, Modi:2022pzm, Dai:2022dso, Qin:2023dew, Jindal:2023qew, Charnock:2019rbk, Doeser:2023yzv, Kostic:2022vok, Bayer:2023rmj}. Indeed, recent studies have shown that field-level information can significantly improve the measurement of cosmological parameters \cite{Leclercq:2021ctr, Andrews:2022nvv, Porqueres:2023drp, beyond2pt, Nguyen:2024yth}. 

A perturbative forward model for field-level analysis and SBI is implemented in the \leftfield\ code \cite{Schmidt:2020ovm} (Lagrangian EFT-based forward model at the field level). In a companion paper \cite{Stadler:2024}, we studied the physical and numerical convergence of the prediction for the rest-frame galaxy density, and we showed that the forward model is sufficiently fast and accurate for the analysis of cosmological volumes. Moving towards the analysis of survey data, the control of observational systematics is as crucial as theoretical uncertainties. Of major concern are redshift-space distortions (RSDs, \cite{1987MNRAS.227....1K,1992ApJ...385L...5H,scoccimarro:2004}), which shift the observed galaxy position $\vecs{x}$ with respect to the rest-frame coordinate $\vec{x}$ by the peculiar velocity $\vec{v}_\tracer$ along the line-of-sight (LOS) direction $\los$:
\begin{equation}
\vecs{x} = \vec{x} + u_{\tracer\pp}\left(\vec{x}\right) \, \los\left(\vec{x}\right)\,,
\quad\mbox{where}\quad
\vec{u}_\tracer \equiv \mathcal{H}^{-1} \vec{v}_\tracer
\label{eq:intro__rsd-displacement}
\end{equation}
is the scaled galaxy velocity, and $u_{\tracer\pp} = \vec{u}_\tracer \cdot \los$ is its projection onto the line of sight. Further, $\mathcal{H} = aH$ denotes the conformal Hubble rate. Redshift space distortions introduce the line of sight as a preferred direction and lead to an anisotropic clustering signal which holds important information on the growth rate of structure $f = d\ln D/d\ln a$ ($D$ is the growth factor and $a$ the cosmological scale factor). They are a vital cosmological probe, but their modeling is complicated by the fact that non-linear effects can have an impact up to relatively large scales.

The field-level forward model of the three-dimensional galaxy density contains all information required to predict the large-scale galaxy velocity field and to boost the density contrast to redshift space. This has been implemented for example in the BORG forward model \cite{Jasche:2018oym,ramanah:2019}. Ref.~\cite{Schmittfull:2020trd} showed that a perturbative field-level modeling of the redshift-space galaxy density exhibits residuals with numerical simulations that are consistent with the noise expectation in the EFT of LSS. In a previous paper \cite{Stadler:2023hea}, we explored the field-level inference of $f$ at fixed initial conditions from N-body halos and found consistent results with percent-level accuracy and precision over a broad range of masses and redshifts. However, this latter work also encountered several challenges in the numerical implementation which raised concerns about the numerical convergence and slowed down the forward model. Here, we address these issues in detail, and we present an implementation of redshift space distortions in \leftfield\ which is significantly faster while providing accurate results with a smaller number of free parameters.  

We start this work with a discussion and validation of the prediction of velocity and momentum fields in section~\ref{sec:velocity} and introduce the redshift-space forward model in section~\ref{sec:forward-model}. We test the model by inferring the growth rate $f$ from N-body halos at the field level, fixing the initial conditions to the ground truth, in section \ref{sec:results}. By fixing the initial conditions,  we remove a considerable source of uncertainty from the parameter constraints, so we can perform very stringent tests. On the other hand, the analysis speeds up considerably such that we can scan through many different analysis configurations. We conclude in section \ref{sec:conclusions}. The appendices contain details on several aspects of the forward model and inference, as well as a range of additional tests.

\section{Validation of momentum and velocity predictions}
\label{sec:velocity}

A key part of the forward model is the gravitational evolution which connects the Gaussian initial conditions $\delta\lin$ to the evolved density contrast $\delta_\matter$. In \leftfield\ \cite{Schmidt:2020ovm}, we only follow modes under perturbative control by applying top-hat filter in Fourier space at the scale $\Lambda$ to the initial conditions, $\delta\lin_\Lambda$ \cite{Schmidt:2020viy}. Gravity is described by Lagrangian Perturbation Theory (LPT), writing the position $\vec{x}$ of any particle in the evolved density field as its initial position $\vec{q}$ plus a displacement $\vec{s}$ that is expanded perturbatively
\begin{equation}
\vec{x}\left(\tau\right) = \vec{q} + \sum_{n=1}^{n_\lpt}\vec{s}^{(n)}\left(\vec{q}, \tau\right)\,,
\label{eq:velocity_eulerian-lagrangian}
\end{equation}
where $\tau$ denotes the conformal time at which the density field is evaluated. The solution for the displacement is governed by the continuity and Poisson equation and can be obtained recursively to any desired order \cite{Rampf:2012up, Zheligovsky:2013eca, Matsubara:2015ipa, Rampf:2015mza, Schmidt:2020ovm}. To this end, it is convenient to decompose the displacement vector into a curl-free (longitudinal) and a curl (transverse) component,
\begin{equation}
\vec{s}^{(n)} \left(\vec{q},\tau\right) = \frac{\vec{\nabla}}{\nabla^2} \sigma^{(n)}\left(\vec{q},\tau\right) - \frac{1}{\nabla^2}\,\vec{\nabla}\times \vec{t}^{(n)}\left(\vec{q},\tau\right)\,.
\label{eq:velocity_displacement-decomposition}
\end{equation}
The linear-order longitudinal component is the Zel'dovich solution $\sigma\lin = -\delta\lin$, and the transverse part starts at third order. In the rest frame, this gravitational evolution can be coupled either with a Lagrangian or with a Eulerian bias expansion.

We describe the numerical implementation of the rest-frame forward model detailed in ref.~\cite{Stadler:2024}. Important for this work, the evolved density $\delta_\matter$ is obtained from the displacement vector by generating an ensemble of uniformly distributed particles with unit weight, shifting each and assigning their density to a grid. Lagrangian bias operators are boosted to the rest frame in a similar manner, by weighting the particles with the operator value at their position. For the displacement of a generic field $\op_{\lagrangian}\left(\vec{q}\right)$ in Lagrangian coordinates this operation yields
\begin{equation}
\left[1 + \delta_\matter\left(\vec{x},\tau\right)\right]\, \op_{\lagrangian}\left(\vec{x},\tau\right)
= \int d^3\vec{q} ~ \op_{\lagrangian}\left(\vec{q}\right)\, \delta_\mathrm{D}\left[ \vec{x} - \vec{q} - \vec{s}\left(\vec{q}, \tau\right)\right]
\,,
\label{eq:velocity_generic-displacement}
\end{equation}
where $1 + \delta_\matter$ is the inverse Jacobian of the coordinate transformation from $\vec{q}$ to $\vec{x}$. The best-practice recommendations for the numerical precision parameters of the rest-frame forward model \cite{Stadler:2024} apply similarly in redshift space and are summarized in figure~\ref{fig:redshiftspace-model__flowchart}. The density assignment step is best computed using a non-uniform fast Fourier transform (NUFFT) algorithm \cite{2019SJSC...41C.479B, 2021A&A...646A..58A}. Rather than the cloud-in-cell (CIC) kernel used in previous work \cite{Stadler:2023hea}, we here exclusively rely on the NUFFT scheme.\footnote{We use the ducc implementation, adapted from \url{https://gitlab.mpcdf.mpg.de/mtr/ducc}.}

\subsection{Velocities in Lagrangian Perturbation Theory}
The peculiar velocity of a non-relativistic particle $\vec{v}\left(\vec{q},\tau\right) = \partial_\tau \vec{s}\left(\vec{q},\tau\right)$ follows readily from the geodesic equation and therefore
\begin{equation}
\vec{u} \left(\vec{q}, \lambda\right) =  f \vec{s}' \left(\vec{q},\lambda\right) = f \sum_{n=1}^{\infty} n\, \vec{s}^{(n)} \left(\vec{q},\lambda\right) \,,
\label{eq:velocity_velocity}
\end{equation}
where a prime denotes derivatives with respect to the time variable $\lambda = \ln D$ and we have approximated the time-evolution of higher-order terms by the Einstein-de Sitter (EdS) solution
\begin{equation}
\sigma^{(n)}\,,~\vec{t}^{(n)} \propto e^{n\lambda} \,.
\end{equation}
Note that the growth factor $D$ as well as the growth rate $f$ are obtained from the $\Lambda$CDM prediction. The EdS approximation has been shown to be accurate to the level of $0.1-0.2\%$ (at $z=0$) in predicting the power spectrum \cite{Schmidt:2020ovm}, while significantly simplifying the numerical computations. From here on, we always assume a fixed time or redshift at which the forward model is evaluated and hence drop the explicit time argument.

In the form of eq.~(\ref{eq:velocity_velocity}), $u_\pp = \vec{u}\cdot\los$ has units of length and directly corresponds to the redshift-space boost due to peculiar velocities. However, it is still evaluated at the Lagrangian coordinate $\vec{q}$, while the boost in eq.~(\ref{eq:intro__rsd-displacement}) depends on $u_\pp\left(\vec{x}\right)$. Applying the particle displacement-assignment operation (eq.~\ref{eq:velocity_generic-displacement}) yields the (LOS projection of the) Eulerian-frame \emph{momentum}
\begin{equation}
\pi_{\pp}\left(\vec{x}\right) = \left[1 + \delta_\matter\left(\vec{x}\right)\right] \, u_{\pp}\left(\vec{x}\right) = \int d^3\vec{q} ~ \delta^\dirac\left(\vec{x} - \vec{q} - \vec{s}\left(\vec{q}\right)\right)\,u_\pp\left(\vec{q}\right)\,.
\label{eq:velocitymodel_eulerian-momentum-density}
\end{equation}
Note that RSDs can also be formulated in terms of $\pi_{\pp}$ \cite{2011JCAP...11..039S}. 
In the previous work \cite{Stadler:2023hea}, we computed eq.~(\ref{eq:velocitymodel_eulerian-momentum-density}) at a high numerical resolution, then divided out the Jacobian $1 + \delta_\matter$ to obtain the velocity in the tracer rest frame, and used this velocity for a second displacement from rest frame to redshift space. We will refer to such a procedure as ``two-step model''. In section~\ref{sec:forward-model}, we introduce an improved model which requires only a single displacement and can operate at a lower numerical resolution. First, however, we investigate the accuracy at which the momentum and the velocity are predicted in \leftfield\ in subsections \ref{sec:velocity_momentum-accuray} and \ref{sec:velocity_velocity-accuray}, respectively.

\subsection{Accuracy of the Eulerian-frame momentum}
\label{sec:velocity_momentum-accuray}

We start with the accuracy of the Eulerian-frame momentum, which follows straightforwardly from displacing the Lagrangian velocity by the vector $\vec{s}$ (eq.~\ref{eq:velocitymodel_eulerian-momentum-density}). Numerical effects on the momentum are very similar to Lagrangian bias operators, which were studied in detail in ref. \cite{Stadler:2024}. We therefore adopt the best-practice recommendations derived there, and we check explicitly that they suppress any resolution effects below the level of the perturbative accuracy in appendix \ref{sec:resolution-effects__momentum}.

\begin{figure}
\centering
\includegraphics[]{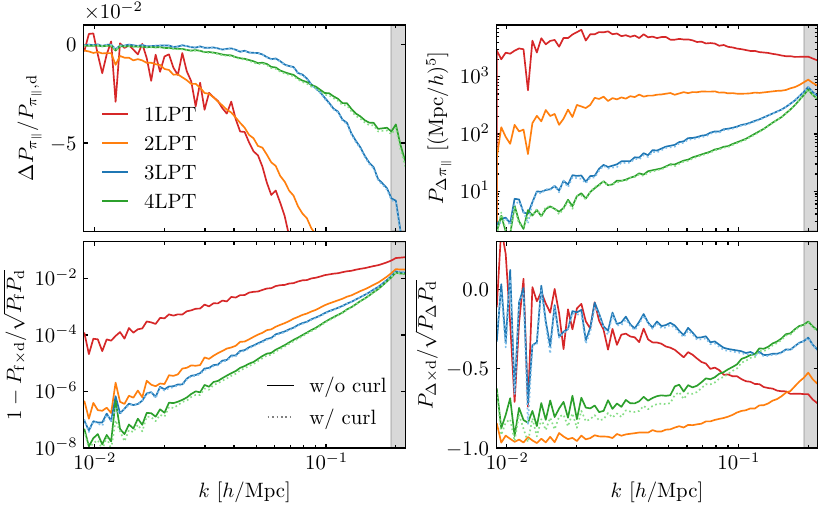}
\caption{Perturbative accuracy of the forward model for the momentum along the LOS-direction, $\pi_\pp$, at $z=0.5$. We compare results from the forward model (subscript ``f'') to data from a N-body simulation which has an identical Fourier-space top-hat filter at $\Lambda=0.20\,h/\mpc$ applied to its initial conditions (subscript ``d''). Clockwise, starting from the top left panel, we show the power spectrum difference, the residual power spectrum, the cross-correlation between forward model and simulation and the cross-correlation between residuals and the simulation. For LPT orders $n_\lpt \geq 3$, we also explore the impact of the transverse displacement component and find it to be negligible. The 4LPT model predicts the momentum power spectrum to better than 4\% accuracy. See figure~\ref{fig:velocity-accuracy__accuracy-velocity-momentum-density_L010_z050} for results at $\Lambda=0.10\,h/\mpc$ and figure~\ref{fig:velocity-accuracy__accuracy-velocity-momentum-density_z100} for $z=1.0$.}
\label{fig:velocity-accuracy__perturbative-accuracy-momentum}
\end{figure}

To evaluate the LPT implementation and its convergence, we compare it to a full N-body simulation run for the same volume and cosmology with an identical cut-off at $\Lambda$ applied to its initial conditions as in the forward model (see ref.~\cite{Stadler:2024} for a detailed description of the reference simulations). The fiducial Euclidean $\Lambda$CDM cosmology assumed throughout this work is identical to refs.~\cite{Stadler:2023hea, Stadler:2024} and defined by
\begin{equation}
\Omega_\mathrm{m} = 0.3\,,
\quad
\Omega_\Lambda = 0.7\,,
\quad
h = 0.7\,,
\quad
\sigma_8 = 0.84\,,
\quad
n_\mathrm{s} = 0.967\,.
\end{equation}
 The comparison between forward model and reference simulation is shown in figure~\ref{fig:velocity-accuracy__perturbative-accuracy-momentum} at $z=0.5$, $\Lambda=0.20\,h/\mpc$, and it is extended to a smaller cut-off, $\Lambda = 0.10\,h/\mpc$, or a higher redshift, $z=1.0$, in the figures~\ref{fig:velocity-accuracy__accuracy-velocity-momentum-density_L010_z050} and \ref{fig:velocity-accuracy__accuracy-velocity-momentum-density_z100}, respectively. As expected, the residual power spectrum (top right) decreases as the order of the LPT solution increases. Moreover, we observe the same alternating behavior for the cross-correlation between residuals and the ground truth (lower right) that we already derived for the rest-frame density \cite{Stadler:2024}. It originates from the fact that the residuals at order $n_\lpt$ are dominated by $\delta^{(n_\lpt+1)}$, and hence their cross-correlation with the ground truth scales at least as $\left\langle \delta^{(2)} \delta^{(2)} \right\rangle \propto k^4$ for odd $n_\lpt$ and at least as $\left\langle \delta^{(1)} \delta^{(3)} \right\rangle \propto k^2 P(k)$ for even ones. At $z=0.5$ and $\Lambda=0.20\,h/\mpc$, the momentum power spectrum is predicted to an accuracy better than 8\% at third and better than 4\% at fourth order LPT, and the cross-correlation coefficient is better than 98.5\%. This means the momentum is somewhat less accurate, by a factor $\sim 4$ in the power spectrum, than the density at identical LPT order. This trend continues at higher redshifts. It can be explained by the prefactor $n$ in \refeq{velocity_velocity}, which enhances the next-order contribution that is no longer included in an $n$-th order forward model by a factor $(n+1)$ relative to the case of the density.  The accuracy of the momentum power spectrum improves considerably as one moves to lower cut-offs, at $\Lambda = 0.10\,h/\mpc$ and $z=0.5$ it is at 2\% and 1\% for the 3LPT and the 4LPT forward model, respectively. Transverse displacement contributions, as figure~\ref{fig:velocity-accuracy__perturbative-accuracy-momentum} illustrates, have a negligible impact.

\subsection{Accuracy of the velocity}
\label{sec:velocity_velocity-accuray}

While the Eulerian-frame momentum can be compared straightforwardly between forward model and simulations, the relevant quantity to predict RSDs is the velocity field. Its computation requires a more complicated volume-weighted assignment rather than a mass-weighted one \cite{Bernardeau:1995en, Bernardeau:1996hb, 1998pcls.work..207V, Pueblas:2008uv, Romano-Diaz:2007eok, Yu:2016mzj, Bel:2018awq, Feldbrugge:2024wcm, Esposito:2024qlo}. We therefore consider two options: (a) divide $\pi_\pp$ in eq.~(\ref{eq:velocitymodel_eulerian-momentum-density}) by the Jacobian prefactor $1+\delta_\matter$ at each grid location; and (b) use a Delaunay tessellation. The Delaunay tessellation divides the space into a set of tetrahedra with a particle at each vertex, and it assigns at each grid point the average velocity over a spherical shell by linearly interpolating over all tetrahedrons which intersect the cell. The radius of the spherical shell and hence the smoothing of the velocity field is set by the size of the Delaunay grid $N_\delaunay$; for more details see appendix~\ref{sec:delaunay}. The Delaunay tessellation is generally too expensive to be routinely applied during a field-level inference. However, for specific realizations it allows to discriminate  errors due to the division by the Jacobian from other numerical and perturbative effects.

\begin{figure}
\centering
\includegraphics[]{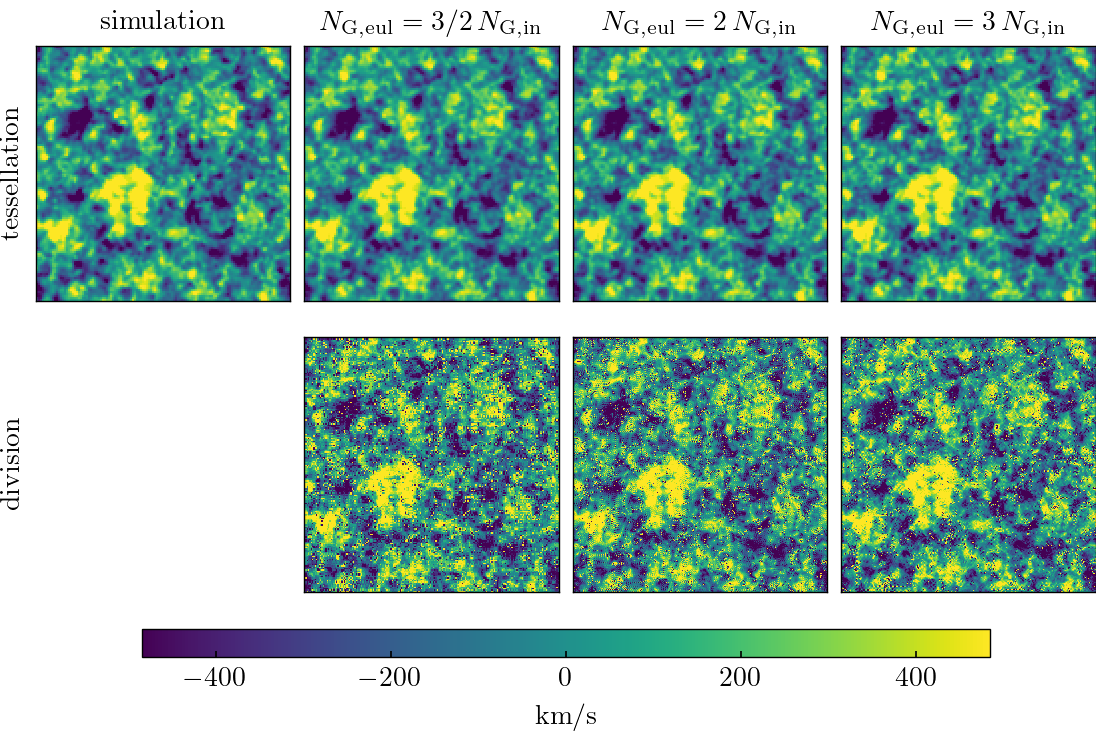}
\caption{Slices through the LOS-velocity $v_\pp$ at $z=0.0$ for a cut-off $\Lambda=0.20\,h/\mpc$, where the LOS direction is perpendicular to the image plane. In the top row, we show velocity grids computed by the Delaunay tessellation for a grid size of $N_\delaunay=128$ from the reference simulation (left) and from the 3LPT forward model at increasing displacement resolution $N_\eul$. We compare these results to velocity grids obtained by dividing out the density from the momentum in the bottom panels. Figure~\ref{fig:velocity-accuracy__presentation__grid-comparison-L010} extends the results to a lower cut-off, $\Lambda=0.10\,h/\mpc$.}
\label{fig:velocity-accuracy__presentation__grid-comparison-L020}
\end{figure}

In figure~\ref{fig:velocity-accuracy__presentation__grid-comparison-L020}, we compare a slice through the Eulerian-frame velocity field between the forward model and the reference simulation which has an identical cut-off applied to its initial conditions. For the forward model, we consider both velocity estimation techniques and in addition vary the resolution of the forward model, $N_\eul$, which also sets the number of particles. While both methods reproduce the overall structures in the velocity field correctly, irregularities in isolated positions from the division by the Jacobian can be noted in the lower panel. In our previous work \cite{Stadler:2023hea}, they were mitigated by the use of relatively large grids, $N_\eul = 384$. However, for a cut-off of $\Lambda = 0.20\,h/\mpc$, computationally acceptable grid sizes cannot fully eliminate the effect. For a smaller cut-off, this situation improves, as figure~\ref{fig:velocity-accuracy__presentation__grid-comparison-L010} shows.

\begin{figure}
\centering
\includegraphics[]{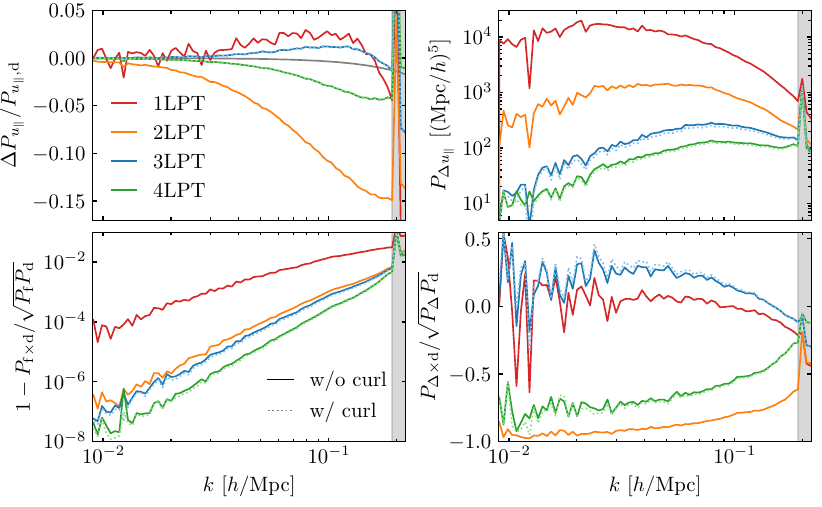}
\caption{Perturbative accuracy of the forward model for the velocity along the LOS-direction, $u_{\pp}$, at $z=0.5$. We compare results from the forward model (subscript ``f'') to data from the reference simulation (subscript ``d''); both use identical initial conditions which are filtered at $\Lambda=0.20\,h/\mpc$. The velocity is computed on a grid of size $N_\delaunay=256$ from simulations with $N_\mathrm{part.}=1536$ and the forward model with $N_\mathrm{part.}=1024$ using the Delaunay tessellation algorithm. Gray lines indicate the expected difference due to the the different number of particles in forward model and simulations (see appendix~\ref{sec:delaunay}), which are subdominant. Clockwise, starting from the top left panel, we show the power spectrum difference, the residual power spectrum, the cross-correlation between forward model and simulation and the cross-correlation between residuals and the simulation. For LPT orders $n_\lpt \geq 3$, we also explore the impact of the transverse displacement component and find it to be negligible. The size of the residuals relative to the reference simulation are comparable for velocities and momentum. See figure~\ref{fig:velocity-accuracy__accuracy-velocity-momentum-density_L010_z050} for results at $\Lambda=0.10\,h/\mpc$ and figure~\ref{fig:velocity-accuracy__accuracy-velocity-momentum-density_z100} for $z=1.0$.}
\label{fig:velocity-accuracy__perturbative-accuracy-velocity}
\end{figure}

Isolated, sharp features in real space impact the power spectrum on all scales in an uncontrollable manner. We therefore focus on the Delaunay tessellation exclusively when quantitatively evaluating the accuracy at which the forward model predicts the velocity. In figure~\ref{fig:velocity-accuracy__perturbative-accuracy-velocity}, we compare the LOS velocity between the forward model and the reference simulations at $z=0.5$ for a cut-off $\Lambda=0.20\,h/\mpc$. In figures~\ref{fig:velocity-accuracy__accuracy-velocity-momentum-density_L010_z050} and~\ref{fig:velocity-accuracy__accuracy-velocity-momentum-density_z100} we extend the comparison to a lower cut-off, $\Lambda=0.10\,h/\mpc$ and a higher redshift, $z=1.0$, respectively. With increasing LPT order the magnitude of the residual power spectrum decreases, as expected. The magnitude of the residuals relative to the ground truth is similar for momentum and velocity. They also scale similarly at low wavenumbers; at high wavenumbers the relative residuals increase slightly less steeply for the velocities compared to momentum (figure~\ref{fig:velocity-accuracy__perturbative-accuracy-momentum}) and density (not shown here). This is in particular the case for $n_\lpt \geq 3$. The power spectrum difference of the velocities oscillates around zero between LPT orders, in particular on intermediate scales. That is, for $n_\lpt = 1,3$ the velocity power spectrum is over-predicted and under-predicted for $n_\lpt=2,4$. In contrast, the density and momentum power spectra are always under-predicted. However, in both cases, the power spectrum differences approach zero from below or above for $n_\lpt$ being even or odd, respectively. Interestingly, the velocity power spectrum is predicted quite accurately at 1LPT order, but the residuals are huge and decrease by more than an order of magnitude when proceeding to second order. This indicates that the good match of $P_{u_\pp}$ at 1LPT is just a coincidence. For the 4LPT model at $z=0.5$, $\Lambda=0.20\,h/\mpc$ the accuracy for the velocity power spectrum is better than $3\%$ and it improves to percent-level accuracy for $\Lambda=0.10\,h/\mpc$ or $z=1.0$. 

While the residual power spectra of momentum and velocity relative to the reference simulation are of similar order, they exceed the corresponding density residuals \cite{Stadler:2024}. As already argued for the momentum, the slower convergence of the LPT solution is likely connected to the additional factor $n$ which multiplies higher-order velocity terms in eq.~(\ref{eq:velocity_velocity}) and is absent in the densities. A deeper investigation of the perturbative convergence of momentum and velocity in Lagrangian Perturbation Theory would be interesting in its own regards but is beyond the scope of this work.

\subsection{Scaling of the leading-order counter terms}

\begin{figure}
\centering
\includegraphics[]{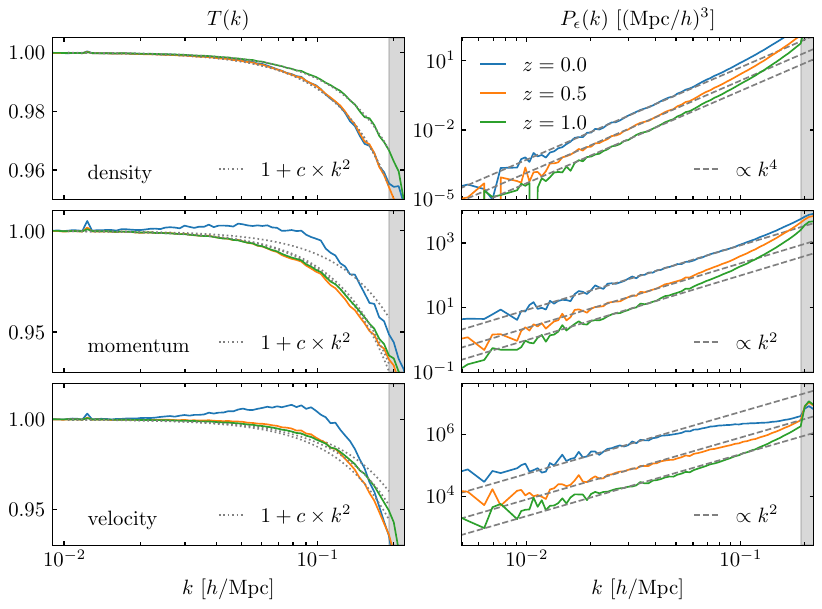}
\caption{Measurement of the deterministic (left) and stochastic (right) counter terms from the comparison of the forward model (cut-off $\Lambda=0.20\,h/\mpc$) with N-body simulations (no initial cut-off). The model predictions are evaluated at 4th order in LPT and include the transverse contribution to the displacement. The bump in the momentum and velocity transfer function is connected to the perturbative accuracy of the forward model and only present for low redshifts and high cut-offs; it vanishes for $\Lambda=0.10\,h/\mpc$. Apart from that, we recover the expected leading-order $k$-scaling for all terms. Note that, in principle, a white noise contribution would be allowed for the velocities. However, it appears to be highly suppressed at least in dark-matter-only simulations.}
	\label{fig:conterterms__summary}
\end{figure}

By comparing the forward model to N-body simulations, now without a cut-off in the initial conditions, we can investigate whether the departure of the forward model from the fully nonlinear matter field matches the EFT expectation for the leading-order counter terms, both stochastic and deterministic. To that end, we express the density, momentum and velocity in the simulations as
\begin{equation}
\alpha_\mathrm{d} = T_\alpha(k)\, \alpha_\mathrm{f} + \epsilon_\alpha 
\quad \mathrm{with} \quad
\alpha = \delta, \pi, u \,,
\end{equation}
where the subscript ``f'' refers to the forward-model quantities which are computed with an explicit cut-off. In contrast, the simulation data (subscript ``d'') has no such cut-off in the linear power spectrum. Note that the comparison is performed in the rest frame and hence the transfer function $T(k)$ does not depend on the angle with the line of sight. It encodes the deterministic part of the bias expansion and should scale as \cite{Baumann:2010tm, Carrasco:2012cv} 
\begin{equation}
T(k) = 1 + c \times k^2 + \ldots \,,
\end{equation}
where ``$\ldots$'' indicates terms of higher order in perturbations and in $k^2$. Stochastic counter terms are accounted for by the noise term $\epsilon$, and we expect the leading order scaling \cite{abolhasani/etal}
\begin{equation}
\left\langle \epsilon_\alpha\left(\vec{k}\right) \epsilon_\alpha\left(\vec{k}'\right) \right\rangle =
 \left(2\pi\right)^3 \delta^\dirac\left(\vec{k} - \vec{k'}\right) P_{\epsilon_\alpha}(k)
 \propto
\begin{cases}
k^2 & \quad \mathrm{for} \quad \alpha = \pi \\
k^4 & \quad \mathrm{for} \quad \alpha = \delta \,, 
\end{cases}
\end{equation}
from momentum conservation (Euler equation) and mass conservation (continuity equation), respectively. On the other hand, for the stochastic velocity component $\epsilon_u$ a white-noise contribution is allowed.

The measurement of the transfer function and the noise power spectrum for density, momentum and velocity is shown in figure~\ref{fig:conterterms__summary}. In particular, we isolate the transfer function $T_\alpha(k)$ from the cross power spectrum between model and simulation $P_{\alpha_\mathrm{d},\alpha_\mathrm{f}}$ and from the model auto power spectrum $P_{\alpha_\mathrm{f},\alpha_\mathrm{f}}$ as 
\begin{equation}
T_\alpha\left(k\right) = P_{\alpha_\mathrm{d},\alpha_\mathrm{f}} / P_{\alpha_\mathrm{f},\alpha_\mathrm{f}} \,.
\end{equation}
The noise power spectrum than follows from the simulation auto power spectrum $P_{\alpha_\mathrm{d},\alpha_\mathrm{d}}$ as
\begin{equation}
P_{\epsilon_\alpha}(k) = P_{\alpha_\mathrm{d},\alpha_\mathrm{d}} - T^2(k) \, P_{\alpha_\mathrm{f},\alpha_\mathrm{f}} \,.
\end{equation}
We indeed observe the expected scaling for the deterministic counter term in figure~\ref{fig:conterterms__summary}. There is a slight bump in the transfer function of momentum and velocity at $z=0$ which is connected to the perturbative accuracy of the forward model at $z=0$. The bump vanishes for lower cut-offs (not shown) and for higher redshifts. It neither is present if we replace the perturbative forward model with N-body simulations with a cut-off in the initial conditions. For the stochastic contribution in density and momentum, we recover the expected leading-order behavior. For the velocities, we find no evidence for a white-noise stochastic component in the dark-matter-only simulations we study here, and $P_{\epsilon_u}$ is found to be consistent with a $k^2$ scaling. Our results in the bottom right panel of figure~\ref{fig:conterterms__summary} show a slight flattening toward low $k$, but larger simulation boxes would be required to assess this trend robustly.

\section{The redshift-space forward model}
\label{sec:forward-model}
The results of section~\ref{sec:velocity_velocity-accuray} demonstrate that the procedure of computing the Eulerian-frame momentum, and dividing by $1+\delta_\matter$ to obtain the velocity (``two-step model'') is prone to numerical inaccuracies. Moreover, it drives up demands on the grid resolution. We therefore reformulate the redshift-space forward model such that only a single displacement operation is necessary, which transforms the density directly from Lagrangian coordinates to redshift space (section~\ref{sec:forward-model__summary}). Subsequently, we investigate the numerical accuracy (section~\ref{sec:forward-model__resolution}), the perturbative accuracy (section~\ref{sec:forward-model__accuracy}) and the computing time (section~\ref{sec:forward-model__timing}) of the forward model.

\subsection{Summary of the redshift-space model}
\label{sec:forward-model__summary}

The gravitational displacement $\vec{s}\left(\vec{q}\right)$ and the Lagrangian-frame velocity $u_\pp\left(\vec{q}\right)$ are given by the LPT solution as described in section~\ref{sec:velocity}. If we take a single displacement step to go from Lagrangian frame directly to redshift space (see below), there is no access to Eulerian-frame quantities. Hence, we focus on the Lagrangian bias expansion
\begin{equation}
\dgd\left(\vec{q}\right) = \sum_{\left\{\op\right\}} b_\op\, \op\left(\vec{q}\right) \,,
\end{equation}
where $\left\{\op\right\}$ includes all scalar invariants up to order $o_\bias$ that can be constructed from the symmetric part of the Lagrangian distortion tensor, $H_{i,j} = \partial_{q_i} s_j$ (see appendix~\ref{sec:bias-operators-density}).
The leading-order linear bias is realized via the operator
\begin{equation}
\sigma^{(1)}\left(\vec{q}\right) = \tr\left[\frac{\partial}{\partial q_i} s^{(1)}_j \left(\vec{q}\right) \right] = - \delta^{(1)}\left(\vec{q}\right) \,.
\label{eq:forward-model__sigma1}
\end{equation}
In general, different orders in the distortion tensors have different time dependencies, and their respective invariants are included with independent coefficients. The exception are the operators $\sigma^{(n)} = \tr M^{(n)}$, which are related to lower-order operators by the LPT recursion relations and hence redundant. This allows us to replace $\sigma^{(1)}$ by $\sigma = \sum_{n=1}^{n_\lpt} \sigma^{(n)}$ for convenient implementation; up to terms beyond $o_\bias$ both formulations are equivalent. We verify that the redefinition only has a very minor impact on the analysis in section~\ref{sec:additional-tests-inference}. 

The tracer velocity field can deviate from $u_\pp$. In the case of galaxies, this bias arises from baryonic effects such as stellar winds and supernova feedback. For halos, spatial averaging and the fact that halos form in special, biased, regions of the density lead to a bias in the velocity \cite{1986ApJ...304...15B, peacock/lumsden/heavens:1987, percival/schaefer:2008, Desjacques:2008jj}. For any type of tracer, however, the equivalence principle ensures that it moves on the same trajectory as matter on large scales. Specifically, bias in the velocities has to be expanded in terms of local observables, which precludes the velocity itself from appearing, and the leading velocity bias contributions are higher-derivative. We therefore express the tracer velocity as
\begin{equation}
\udgpp\left(\vec{q}\right) = u_\pp\left(\vec{q}\right) + \sum_{\{\uop\}} \beta_\uop\, \uop\left(\vec{q}\right)\,,
\label{eq:forward-model__velocity-bias}
\end{equation}
where $\{\uop\}$ is a set of higher-order velocity bias operators constructed by contracting the distortion tensor with spatial derivatives and the LOS direction \cite{Stadler:2023hea}, see appendix \ref{sec:bias-operators-velocity}. The maximum order of the velocity expansion is denoted by $o_{u,\bias}$.

The redshift-space density follows from first displacing $\dgd\left(\vec{q}\right)$ by $\vec{s}\left(\vec{q}\right)$ and then by $\udgpp\left(\vec{x}\right)$. This operation can be written as a single displacement (``one-step model'')
\begin{equation}
\vec{q} \longrightarrow \vecs{x}\left(\vec{q}\right) = \vec{q} + \vec{s}\left(\vec{q}\right) + \udgpp\left(\vec{q}\right) \los \,.
\end{equation}
In particular, the Jacobian of the mapping from $\vec{q}$ to $\vecs{x}$ is
\begin{equation}
\left. \left|\frac{\partial^3 \vecs{x}}{\partial^3\vec{x}}\right|^{-1}\,\left|\frac{\partial^3 \vec{x}}{\partial^3\vec{q}}\right|^{-1} \right._{\substack{\vec{q} = \vec{q}\left(\vec{x}\right) \\ \vec{x}=\vec{x}\left(\vecs{x}\right)}} ~
=
\left|\frac{\partial^3\vecs{x}}{\partial^3 \vec{q}}\right|^{-1}_{\vec{q}=\vec{q}\left(\vecs{x}\right)} 
= \frac{1 + \delta_\matter\left(\vecs{x}\right)}{1 + \partial_{x_\pp} \udgpp\left(\vecs{x}\right)}\,.
\end{equation}
For example, the displacement of a uniform field, and using the matter velocity, yields the familiar redshift-space density
\begin{equation}
1+\tilde{\delta}_\matter(\vecs{x})
= \frac{1 + \delta_\matter\left(\vecs{x}\right)}{1 + \partial_{x_\pp} u_\parallel\left(\vecs{x}\right)} \,.
\end{equation}
For the biased tracer field, the redshift-space displacement becomes 
\begin{align}
1 + \dgds\left(\vecs{x}\right) &= \int d^3\vec{q}~\delta^\dirac \left[\vecs{x} - \vec{q} - \vec{s}\left(\vec{q}\right) - \udgpp\left(\vec{q}\right)\los\right] \,\left[1 + \dgd\left(\vec{q}\right)\right] \nonumber \\
&= \frac{1 + \delta_\matter\left(\vec{x}\left(\vecs{x}\right)\right)}{1 + \partial_{x_\pp} \udgpp \left(\vec{x}\left(\vecs{x}\right)\right)} + \frac{\sum\nolimits_{\left\{\op\right\}} \left[1 + \delta_\matter\left(\vec{x}\left(\vecs{x}\right)\right)\right]\, b_\op\, \op\left[\vec{q}\left(\vecs{x}\right)\right]}{1 + \partial_{x_\pp} \udgpp\left(\vec{x}\left(\vecs{x}\right)\right)} \,,
\label{eq:forward-model__bias-expansion-zspace}
\end{align}
The Jacobian factor of the displacement to redshift space $\left(1 + \partial_{x_\pp}\udgpp\right)^{-1}$ ensures the correct redshift-space density. As in the rest-frame implementation of Lagrangian bias, the bias operators are multiplied by an additional Jacobian factor $1+\delta_\matter$ from the Lagrangian-to-rest frame mapping. At leading order, they correspond to the desired Lagrangian operators shifted to redshift space $\mathcal{O}\left[\vec{q}\left(\vecs{x}\right)\right]$, and at higher orders the multiplication by $\delta_\matter$ can be absorbed by a redefinition of the bias coefficients (note that this prefactor is the rest-frame matter density evaluated at the redshift-space position). In the analysis, all bias coefficients are varied separately, and to implement this efficiently we compute the displacements of all terms in eq.~(\ref{eq:forward-model__bias-expansion-zspace}) independently. In this way, we can scan through different bias coefficients (at fixed cosmology) more efficiently without the need to reevaluate the displacement at every new value tested.

In comparison to the real-space forward model in \leftfield, the redshift-space formulation gives no explicit access to Eulerian-frame quantities. This has several ramifications for the bias expansion. While the linear bias term previously was expressed as $b_1 \delta_\matter$ \cite{Schmidt:2020ovm, Stadler:2024} it now becomes
\begin{equation}
  \tilde{\delta}_\matter \left(\vecs{x} \right) + \frac{b_\sigma \,\sigma\left(\vecs{x}\right)}{1 + \partial_{x_\pp} \udgpp \left(\vecs{x}\right)}
= \frac{1+ \delta_\matter(\vecs{x}) + b_\sigma \,\sigma\left(\vecs{x}\right)}{1 + \partial_{x_\pp} u_\parallel \left(\vecs{x}\right)} -1 \,.
\label{eq:forward-model__linear-bias}
\end{equation}
Both formulations can be transformed into each other via a redefinition of the bias coefficients, and at linear order this relation is evident from the definition of $\sigma^{(1)}$ in eq.~(\ref{eq:forward-model__sigma1}). Hence we expect the two formulations to yield equivalent results.

In the rest-frame forward model, it is most convenient to construct higher-order derivative operators in Eulerian coordinates to save the numerical costs of their displacement. In the redshift-space forward model, we now construct higher-derivative operators as well as the velocity bias operators in Lagrangian coordinates. The difference between Lagrangian and Eulerian derivatives,
\begin{equation}
\frac{\partial}{\partial x_j} = \left[\delta_{ij} + H_{ij}\left(\vec{q}\right)\right]^{-1} \frac{\partial}{\partial q_i}
\label{eq:forward-model__eulerian-derivatives}
\end{equation}
can be absorbed to any fixed order by a redefinition of the bias coefficients $b_\op$. Similarly, Lagrangian and Eulerian derivatives yield an equivalent basis for the velocity bias operators up to next-to-leading order, as is shown explicitly in appendix~\ref{sec:bias-operators-velocity}.

We provide a detailed comparison between the one-step and the two-step implementations of the redshift-space forward model in appendix~\ref{sec:two-step-comparison}. The present implementation is similar to the redshift-space forward model of ref.~\cite{Schmittfull:2020trd} in that it computes a single displacement step from Lagrangian coordinates to redshift space; in contrast, the field-level forward model of refs.~\cite{Jasche:2018oym,ramanah:2019,Lavaux:2019fjr} first computes the matter density field in redshift space, and then applies a local bias expansion on this field. 
Unlike ref. \cite{Schmittfull:2020trd}, our forward model performs the exact (unexpanded) displacement operation using the shift vector into redshift space evaluated up to order $n_\lpt$. In doing so, we account for shift terms which are protected by symmetries to arbitrarily high order, and the computation of bias and RSD is done in the correct sequence.

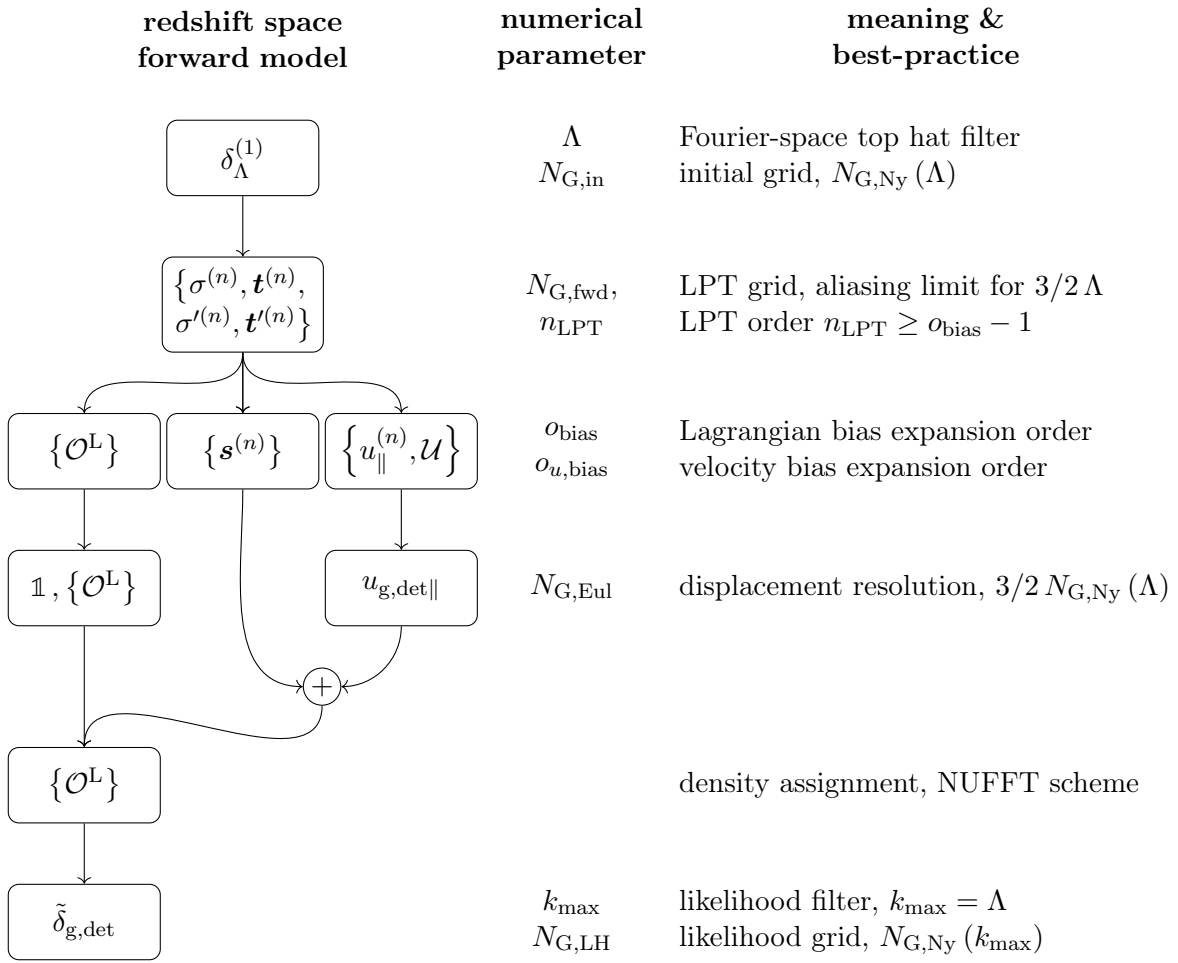
\begin{figure}
\begin{tikzpicture}[remember picture]

\newlength{\hmargin}
\setlength{\hmargin}{1in+\hoffset+\oddsidemargin}

\newlength{\nodeshiftV}
\setlength{\nodeshiftV}{.8cm}

\newlength{\nodeheight}
\setlength{\nodeheight}{1cm}

\newlength{\nodewidth}
\setlength{\nodewidth}{2.cm}

\newlength{\nodeshiftH}
\setlength{\nodeshiftH}{0.2\textwidth}

\newlength{\bbshift}
\setlength{\bbshift}{0.28\textwidth}


\tikzset{leftfieldnode/.style={anchor=north, draw, rounded corners, minimum height=\nodeheight, minimum width=\nodewidth, align=center}};

\tikzset{bbnode/.style={anchor=east, text width=0.42\textwidth, align=left, xshift=\linewidth}};

\node[leftfieldnode] at (current page.north west) (deltaL) {$\delta\lin_\Lambda$};

\node[leftfieldnode, yshift=-\nodeshiftV] at (deltaL.south) (sigman) {$\left\{ \sigma^{(n)}, \vec{t}^{(n)}, \right.$\\ $\left. \sigma'^{(n)}, \vec{t}'^{(n)} \right\}$};

\node[leftfieldnode, yshift=-\nodeshiftV] at (sigman.south) (sn) {$\left\{ \vec{s}^{(n)} \right\}$};

\node[leftfieldnode, anchor=north west, xshift=-\nodeshiftH] at (sn.north) (biasL) {$\left\{\op^\lagrangian\right\}$};

\node[leftfieldnode, anchor=north east, xshift=\nodeshiftH] at (sn.north) (upp) {$\left\{ u^{(n)}_\pp, \uop \right\}$};

\node[leftfieldnode, anchor=north, yshift=-\nodeshiftV] at (upp.south) (udgpp) {$\udgpp$};

\draw[opacity=0] ([yshift=-\nodeshiftV] udgpp.south) to[] node[opacity=1, midway, circle, anchor=center, draw, inner sep=1pt] (add) {$+$} ([yshift=-\nodeshiftV] sn.south|-udgpp.south);

\node[leftfieldnode, yshift=-\nodeshiftV] at (biasL.south) (dispopset) {$\mathbb{1}\,,\left\{\op^\lagrangian\right\}$};

\node[leftfieldnode, yshift=-2\nodeshiftV] at (dispopset.south) (opsE) {$\left\{\op^\lagrangian\right\}$};

\node[leftfieldnode, yshift=-\nodeshiftV] at (opsE.south) (dgds) {$\dgds$};


\node[anchor=center, xshift=\bbshift, text width=3cm, align=center] at (deltaL.center) (deltaLparam) {$\Lambda$\\ $N_\ini$};
\node[anchor=center, xshift=\bbshift, text width=3cm, align=center] at (sigman.center) {$N_\fwd$,\\$n_\lpt$};
\node[anchor=center, xshift=\bbshift, text width=3cm, align=center] at (sn.center) {$o_\bias$ \\ $o_{u,\bias}$};
\node[anchor=center, xshift=\bbshift, text width=3cm, align=center] at (deltaL.center|-udgpp.center) {$N_\eul$};
\node[anchor=center, xshift=\bbshift, text width=3cm, align=center] at (deltaL.center|-dgds.center) {$k_\mathrm{max}$\\ $N_\lh$};

\node[bbnode] at (biasL.west|-deltaL.center) {
Fourier-space top hat filter\\	
initial grid, $N_{\grid,\nyquist}\left(\Lambda\right)$};
\node[bbnode] at (biasL.west|-sigman.center) {
LPT grid, aliasing limit for $3/2\,\Lambda$\\
LPT order $n_\lpt \geq o_\mathrm{bias}-1$};
\node[bbnode] at (biasL.west|-sn.center) {
Lagrangian bias expansion order \\
velocity bias expansion order
};
\node[bbnode] at (biasL.west|-udgpp.center) {
displacement resolution,
$3/2\, N_{\grid,\nyquist}\left(\Lambda\right)$};

\node[bbnode] at (biasL.west|-opsE.center) {density assignment, NUFFT scheme};

\node[bbnode] at (biasL.west|-dgds.center) {likelihood filter, $k_\mathrm{max}=\Lambda$\\
likelihood grid, $N_{\grid,\nyquist}\left(k_\mathrm{max}\right)$
};

\node[anchor=south, yshift=\baselineskip, text width=3cm, align=center] at (deltaLparam.north) (headerP) {\textbf{numerical\\ parameter}};
\node[anchor=center, text width=5cm, align=center] at  (deltaL.center|-headerP.center) {\textbf{redshift space\\ forward model}};
\node[bbnode, align=center] at  (biasL.west|-headerP.center) {\textbf{meaning \&\\ best-practice}};

\draw[->] (deltaL.south) -- (sigman.north);
\draw[->] (sigman.south) -- (sn.north);
\draw[->] (sigman.south) -- (sn.north);
\draw[->] (biasL.south) -- (dispopset.north);
\draw[->] (sigman.south) to[out=-90, in=90] (biasL.north);
\draw[->] (sigman.south) to[out=-90, in=90] (upp.north);
\draw[->] (dispopset.south) -- (opsE.north);
\draw[->] (opsE.south) -- (dgds.north);
\draw[->] (udgpp.south) to[out=-90, in=0] (add.east);
\draw[->] (sn.south) to[out=-90, in=180] (add.west);
\draw[->] (add.south) to[out=-90, in=90] (opsE.north);
\draw[->] (upp.south) -- (udgpp.north);
\end{tikzpicture}
\caption{Forward model of the galaxy density field $\dgds$ in redshift space. The model requires only a single displacement from the Lagrangian frame to redshift space by adding up the LPT displacement vector and the tracer velocity $\udgpp$. The latter includes higher-order bias operators $\{\uop\}$. The density bias operators $\{\op_{\lagrangian}\}$ are constructed in the Lagrangian frame and then displaced to redshift space. In the middle column, we list the parameters which control the numerical accuracy of the forward model. Explanations and best-practice recommendations are given in the right column (see also ref.~\cite{Stadler:2024}).}
	\label{fig:redshiftspace-model__flowchart}
\end{figure}

For the likelihood, we follow the same approach as in ref.~\cite{Stadler:2023hea}, and formulate it in Fourier space
\begin{equation}
\ln \mathcal{P}\left(\tilde{\delta}_\tracer | \dgds\right) = -\frac{1}{2} \int_{|\vec{k}| < k_\mathrm{max}} \frac{d^3\vec{k}}{\left(2\pi\right)^3} \, \left\{ \frac{\left| \tilde{\delta}_\tracer - \dgds\right|^2}{P_\epsilon\left(\vec{k}\right)} + \ln \left[2\pi P_\epsilon\left(\vec{k}\right) \right]\right\} \,.
\label{eq:forward-model__likelihood}
\end{equation}
We allow for the leading-order white noise contribution as well as two scale-dependent components. One of the scale-dependent noise terms is isotropic, the other depends on the cosine of the LOS-angle $\mu$,
\begin{equation}
P_\epsilon\left(\vec{k}\right) = P_{\epsilon,0} \left[ 1 + \sigma_{\epsilon,2} k^2 + \sigma_{\epsilon\mu,2} \left(\mu k\right)^2\right] \,.
\label{eq:forward-model__noise}
\end{equation}
The anisotropic term accounts for the leading-order impact of noise in the velocities. In this form, the likelihood neglects the transformation of isotropic noise in the rest-frame galaxy density $\epsilon_\tracer\left(\vec{x}\right)$ to redshift space \cite{Cabass:2020jqo}
\begin{equation}
\tilde{\epsilon}_\tracer\left(\vecs{x}\right) = \left. \frac{\epsilon_\tracer\left(\vec{x}\right)}{1 + \partial_{x_\pp} u_\pp\left(\vec{x}\right)} \right|_{\vec{x} = \vec{x}\left(\vecs{x}\right)} \,.
\label{eq:forward-model__rsd-noise}
\end{equation}
Parameter inferences at fixed initial conditions were observed to be rather insensitive to the precise form assumed for the likelihood. Indeed, we verify this explicitly for the growth rate by comparing two data sets generated from matter particles in N-body simulations. In the first case, we subsample the particles in the rest frame before displacing them to redshift space to generate anisotropic noise according to eq.~(\ref{eq:forward-model__rsd-noise}). For the second dataset, we displace all matter particles and add white noise of equivalent power. The analysis of both data sets with the likelihood of eq.~(\ref{eq:forward-model__likelihood}) yields consistent results. However, we plan to revisit the noise modeling and its impact on the inference in future work to develop a model that can properly account for scale-dependent noise and the transformation from rest frame to redshift space.

The computational steps of the redshift-space forward model are summarized in figure~\ref{fig:redshiftspace-model__flowchart}, where we also list the parameters governing the numerical accuracy. The computation starts from $\delta\lin_\Lambda$, the linear density filtered at the cut-off scale $\Lambda$, and leads to the predicted redshift-space tracer field, $\dgds$. We turn to the numerical and perturbative accuracy in the following sections.

\subsection{Resolution effects}
\label{sec:forward-model__resolution}

Due to the initial cut-off and the perturbative nature of the forward model, it is generally possible to predict the highest wavenumber with non-zero excitation. By choosing grids large enough that the Nyquist frequency $k_\nyquist$ exceeds this highest wavenumber, resolution effects can be eliminated. The minimum grid size to fulfill this condition is
\begin{equation}
N_{\grid,\nyquist}\left(k\right) = \left\lceil \frac{k\, L_\mathrm{box}}{\pi} \right\rceil\,,
\end{equation}
where $L_\mathrm{box}$ is the side length of a cubic simulation box and $\lceil\ldots\rceil$ indicates up-rounding to the next integer. One the other hand, if a field has support up to $n\Lambda$ and we are only interested in modes below $\Lambda$, it is sufficient to represent the field on a grid large enough to avoid aliasing of high-$k$ modes into the modes of interest. The corresponding minimum grid size is $N_{\grid,\nyquist}\left(k_\mathrm{alias}\right)$ with
\begin{equation}
k_\mathrm{alias} = \frac{n+1}{2}\, \Lambda\,.
\end{equation}
The exception from these simple relations is the displacement; as a completely non-linear operation it can populate modes up to arbitrarily high $k$. A good criterion to suppress resolution effects in the displacement is choosing $N_\eul = 3/2\, N_{\grid,\nyquist}(\Lambda)$, which prevents aliasing from quadratic interactions of linearly populated modes \cite{Stadler:2024, 1971JAtS...28.1074O}. For the forward model with a one-step displacement, the same theoretical considerations apply as for the rest-frame model. Higher-order terms which are neglected due to the finite grid resolution scale at least as $k^2$ and pick up another factor of $k$ from the convolution of top-hat filters in the initial conditions. In the simplest case where $N_\eul = N_{\grid,\nyquist}\left(\Lambda\right)$, residuals due to resolution effects therefore can be shown to scale as \cite{Stadler:2024}
\begin{equation}
\left\langle \Delta\tilde{\delta}_\mathrm{res.}\left(\vec{k}\right) \Delta\tilde{\delta}_\mathrm{res.}\left(\vec{k}'\right) \right\rangle^\LO
= \left(2\pi\right)^3 \delta^\dirac\left(\vec{k} + \vec{k'} \right) \,P_{\Delta\tilde{\delta},\mathrm{res.}}
 \propto k^6\,.
\label{eq:redshiftspace-model__resolution-residuals}
\end{equation}

\begin{figure}
\centering
\includegraphics[]{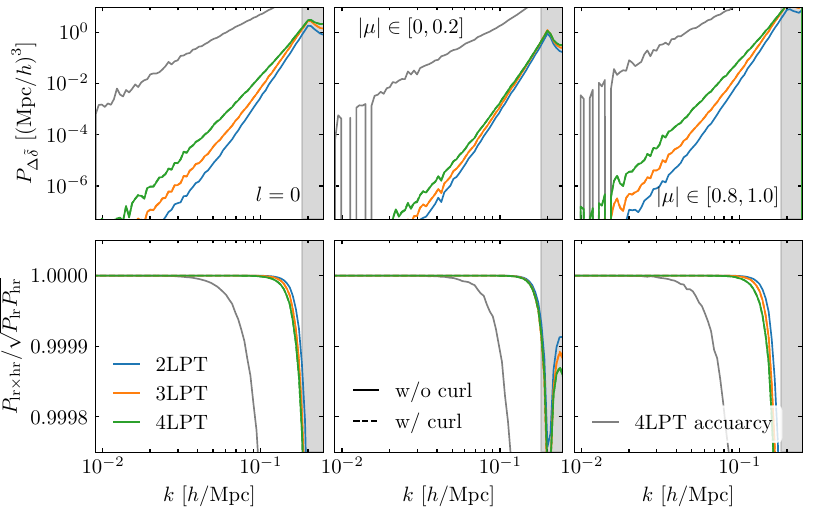}
\caption{Resolution effects in the redshift-space forward model. We consider the one-step model for $\Lambda=0.20\,h/\mpc$, $z=0.0$ and compare results at different LPT orders for $N_\eul=3/2\, N_{\grid,\nyquist}(\Lambda) = 192$ (subscript ``lr'') to a high-resolution reference with $N_\eul=1024$ (subscript ``hr''). In the top row, we show the residual power spectrum and in the bottom one the cross-correlation with the high-resolution reference; differences are presented in terms of the power spectrum monopole (left) and two wedges perpendicular (middle) and parallel (right) to the line of sight, where $\mu$ is the cosine between $\vec{k}$ and the LOS direction. As expected, resolution effects are stronger at higher LPT orders, but the increase is not very severe. Transverse contributions to the displacement have negligible impact on the numerical accuracy. For reference, the gray line shows the perturbative error of the 4LPT model at $z=0$ (see section~\ref{sec:forward-model__accuracy}), which dominates over the resolution residuals. The results are extended to different resolutions $N_\eul$ for the 3LPT model in figure~\ref{fig:redshiftspace-accuracy__resolution-3LPT-summary}.}
\label{fig:redshiftspace-accuracy__resolution-nLPT-summary}
\end{figure}

Resolution effects in the redshift-space 3LPT model for varying $N_\eul$ are depicted in figure~\ref{fig:redshiftspace-accuracy__resolution-3LPT-summary}, where we compare the forward model to a high-resolution reference. Indeed, we recover the expected $k^6$ scaling, and we note a rapid decrease of the residuals with increasing resolution. In figure~\ref{fig:redshiftspace-accuracy__resolution-nLPT-summary}, we show the residuals due to resolution effects for $N_\eul=3/2\, N_{\grid,\nyquist}\left(\Lambda\right)$ at different LPT orders and compare them to the perturbative residuals (see section~\ref{sec:forward-model__accuracy}). Resolution effects are larger for higher LPT orders, which populate higher modes in $\vec{s}$ and $u_\pp$, but the effect is not very severe. Further, they are largest parallel to the line of sight where the velocity contribution has the largest impact. Importantly, the 3/2-rule is sufficient to suppress resolution effects well below the perturbative accuracy. This is a significant improvement over the previous two-step implementation which required larger grid sizes and correspondingly longer computation times (see appendix~\ref{sec:two-step-comparison}).

\subsection{Perturbative accuracy}
\label{sec:forward-model__accuracy}

\begin{figure}
\centering
\includegraphics[]{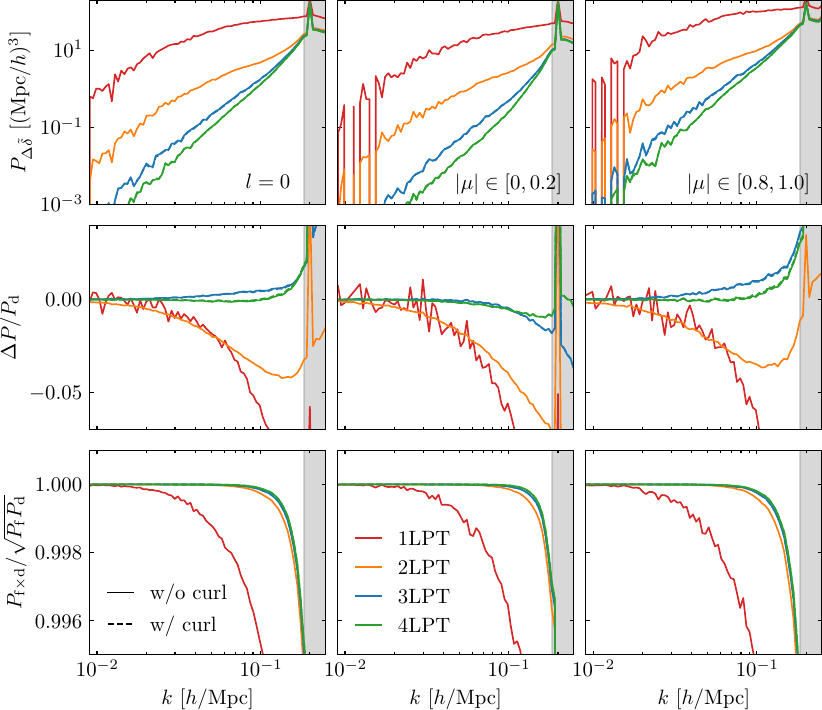}
\caption{The perturbative error of the redshift-space forward model at $z=0.5$. We compare the forward model at different LPT orders (subscript ``f'') to reference data from a N-body simulation (subscript ``d''); both start from identical initial conditions with a cut-off at $\Lambda=0.20\,h/\mpc$. Shown are the residual power spectrum (top), the difference in power spectra  (middle) and the cross-correlation between model and simulation (bottom) in terms of the power spectrum monopole (left) and two power spectrum wedges perpendicular (middle) and parallel (right) to the line of sight ($\mu$ is the cosine between $\vec{k}$ and the LOS direction). For LPT orders $n_\lpt \geq 3$, we distinguish between results with and without the transverse displacement contribution; however, the impact is negligible. The corresponding results for a lower cut-off are shown in figure~\ref{fig:redshiftspace-accuracy__lpterror_L010_z050}.}
\label{fig:redshiftspace-accuracy__lpterror_z050}
\end{figure}

To evaluate the accuracy of the forward model, we compare it to the reference simulations which have an identical cut-off in their initial conditions. The results are shown in figure~\ref{fig:redshiftspace-accuracy__lpterror_z050}. As expected, the residual power spectrum continuously decreases as the LPT order increases. For $\Lambda=0.20\,h/\mpc$ and $z=0.5$, the power spectrum and the cross-correlation with the reference is accurate at the level of 4\% for the 2LPT model, and the power spectrum accuracy further improves to 2\% for higher LPT orders. At lower cut-off values, the accuracy improves and the power spectrum is predicted at the sub-percent level for $\Lambda=0.10\,h/\mpc$, $z=0.5$ and $n_\lpt \geq 3$ (see figure~\ref{fig:redshiftspace-accuracy__lpterror_L010_z050}). The transverse contribution to the displacement vector has negligible impact on the forward model accuracy, similarly to the rest frame \cite{Stadler:2024}.

\begin{figure}
\centering
\includegraphics[]{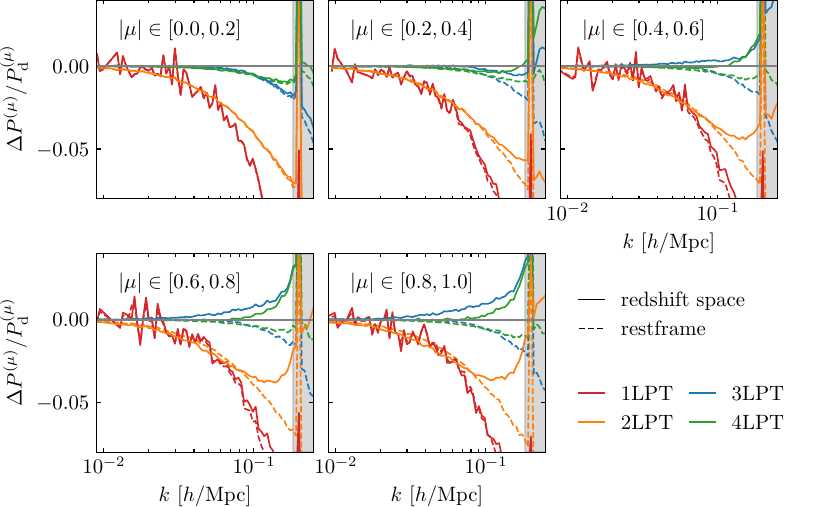}
\caption{Power spectrum accuracy for the redshift-space forward model (solid lines) in comparison to the rest frame (dashed lines) for five wedges, where $\mu$ is the cosine between $\vec{k}$ and the LOS direction. We compare the forward model predictions to a N-body simulation (subscript ``s'') which has identical initial conditions with a cut-off at $\Lambda=0.20\,h/\mpc$ for a redshift $z=0.5$.}
\label{fig:redshiftspace-accuracy__lpterror-restframecomparison_z050}
\end{figure}

The accuracy of the redshift-space forward model is compared to the rest-frame results \cite{Stadler:2024} in terms of five $k$-wedges in figure~\ref{fig:redshiftspace-accuracy__lpterror-restframecomparison_z050}. Along the line-of-sight, the two show excellent agreement as one would expect. While the rest-frame power spectrum is always under-predicted at high wavenumbers, the redshift-space forward model can exceed the ground truth in particular along the LOS direction where velocities are the most important. Indeed, we found in section~\ref{sec:velocity_velocity-accuray} that the velocity power spectrum can be over-predicted by the forward model, and in the absence of shell-crossing this is expected to enhance the clustering along the line-of-sight direction. However, it is not straightforward to relate the velocity and density auto-correlation in the rest frame to the redshift-space power spectrum, which also contains cross-correlations between density, velocity and higher-order velocity moments \cite{2011JCAP...11..039S}. Terms at higher orders in $(k\mu)$ enter with alternating signs and become increasingly important for high wavenumbers closely aligned with the line of sight, explaining the boost in power and its dependence on the LOS direction.

In general, errors in the velocities become increasingly important parallel to the LOS and increase the perturbative error there. An interesting exception is the 2LPT model, where errors in the velocity and the displacement apparently cancel each other and the power spectrum accuracy is better along the line-of-sight than perpendicular to it. However, in this case as in all others, the power spectrum of the residuals (not shown) is always smallest perpendicular to the line of sight.

\subsection{Computing time}
\label{sec:forward-model__timing}

\begin{figure}
\centering
\includegraphics[]{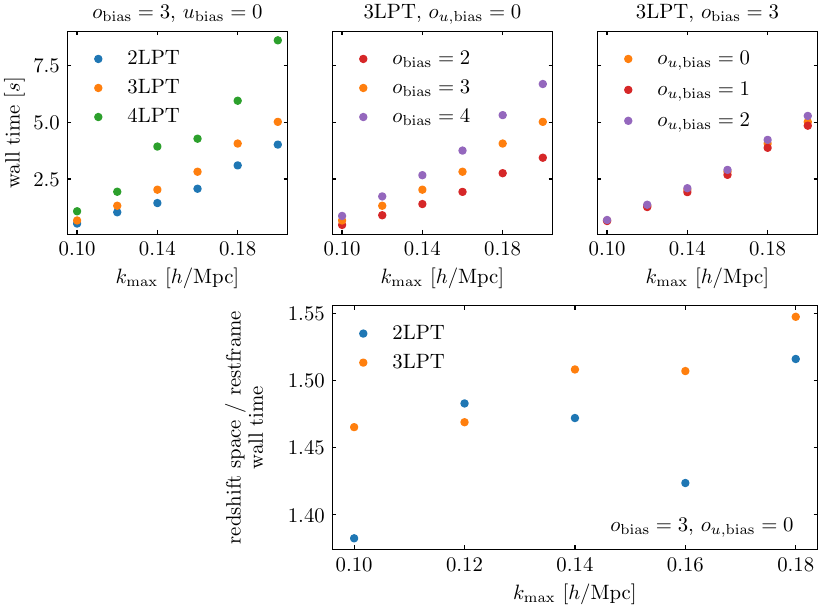}
\caption{Average wall time over ten evaluations of the forward model run on a Intel(R) Xeon(R) Gold 6138 CPU \@ 2.00GHz with 20 cores. The box size is $8\,\left(\mathrm{Gpc}/h\right)^3$ , and we compare different cut-off values and forward model choices. Precision settings not explicitly specified are set to the best-practice recommendations (see figure~\ref{fig:redshiftspace-model__flowchart}). In the bottom panel, we compare the evaluation time between the redshift-space and the rest-frame forward model, where the latter is taken from ref.~\cite{Stadler:2024}. The redshift-space model displaces a larger number of operators, partially explaining the increase in runtime. Overall, a single evaluation of the redshift-space model takes half a second to a few seconds.}
\label{fig:timing}
\end{figure}

For field-level analysis and for SBI, a fast evaluation of the forward model is crucial, in particular to cover volumes of current-generation galaxy surveys. Here, we focus on a $8\,\left(\mathrm{Gpc}/h\right)^3$ cubic simulation box, and we explore the execution time of the redshift-space forward model on a Intel(R) Xeon(R) Gold 6138 CPU \@ 2.00GHz with 20 cores in figure~\ref{fig:timing}. The identical architecture was used to test the rest-frame model \cite{Stadler:2024}, to facilitate direct comparison.

Overall, it is possible to evaluate the redshift-space galaxy density within half a second to a few seconds. The execution time clearly increases with the cut-off scale $\Lambda$ and with LPT order. Each requires larger grids $N_\fwd$ for the computation of the LPT solution; the former in addition demands an increase in $N_\eul$ while the latter necessitates additional operations to find the higher-order LPT terms. While the density assignment always dominates the execution time, its contribution reduces from $(51 - 57)\,\%$ to $(32 - 39)\,\%$ between the 3LPT and the 4LPT model. Correspondingly, the time spent in obtaining the LPT computations increases from $(11 - 16)\,\%$ to $(29 - 33)\,\%$.

With the density assignment dominating the execution time, we see a clear dependence on the order of the bias expansion $o_\mathrm{bias}$. A higher bias order implies a larger number of operators to be displaced, in the case illustrated in the top middle panel of figure~\ref{fig:timing}, we have 5, 9 and 12 operators for $o_\mathrm{bias} = 2,3,4$, respectively. These numbers are higher than for the corresponding rest-frame scenarios, firstly because we now construct higher derivative operators in the Lagrangian rather than the Eulerian frame, and second due to the implementation of the linear bias relation in redshift space (eq.~\ref{eq:forward-model__linear-bias}). 

In contrast to the density operators, increasing the order of the velocity bias expansion has a much lesser impact on the computing time. The number of velocity operators is 0, 1 and 4 for $u_\mathrm{bias}=0,1,2$ considered in the top right panel of figure~\ref{fig:timing}. This is because $\udgpp$ (eq.~\ref{eq:forward-model__velocity-bias}) enters the displacement directly and the velocity operators themselves are not boosted. However, in the sampling of velocity bias coefficients, each new value tried for $\beta_\uop$ now necessitates to recompute the full displacement operation. In contrast, the density coefficients can be varied very efficiently and even be marginalized over analytically \cite{Elsner:2019rql}.

In summary, the redshift-space model is more costly than its restframe equivalent only by a factor of $\sim 1.5$ that is largely independent of the cutoff. This suggests that field-level and simulation-based inferences of cosmological parameters such as shown for the rest-frame case in ref.~\cite{Nguyen:2024yth} are already feasible in redshift space.

\section{Growth-rate inference at fixed initial conditions}
\label{sec:results}

The forward model residuals (section~\ref{sec:forward-model__accuracy}) are indicative of the accuracy of the model, but are not easily translated quantitatively into the accuracy and precision that we can expect in a field-level parameter inference. To this end, we perform a set of parameter recovery tests from halos in redshift space in N-body simulations. The data set and analysis setup are detailed in section~\ref{sec:results_analysis}, we present the results in section~\ref{sec:results_results} and put the precision of our tests in context in section~\ref{sec:results_precision}.

\subsection{Analysis}
\label{sec:results_analysis}
We test the forward model by inferring the growth rate $f$ from N-body halos in redshift space. The halos provide a non-trivially biased tracer of the underlying non-linear matter density. We have two simulation realizations available, for which we consider three redshift snapshots at $z=0.0,\,0.5,\,1.0$. Halos are identified in their rest frame with the ROCKSTAR halo finder \cite{Behroozi2013}, in contrast to previous redshift-space analyses with \leftfield\ that used the same simulations but focused on AMIGA halos \cite{2009ApJS..182..608K}. In particular we consider two halo mass bins, $\lg M/\left(h^{-1} M_\odot\right) \in \left[12.5, 13.0\right]$ and $\lg M/\left(h^{-1} M_\odot\right) \in \left[13.0, 13.5 \right]$. To boost the halos to redshift space, we displace their centers by the center-of-mass velocity of the halo \emph{core}, i.e. the mass-averaged velocity within 10\% of the halo radius. We provide a detailed comparison with results from the two-step model \cite{Stadler:2023hea} on this data set in appendix~\ref{sec:two-step-comparison-inference}. For further details on the simulations see ref.~\cite{Stadler:2024} and appendix F therein. 

To derive constraints on the growth rate $f$, we marginalize over the noise amplitudes in eq.~(\ref{eq:forward-model__noise}) and over the density and velocity bias coefficients $b_\op$ and $\beta_\uop$. The coefficients $b_\op$ enter the likelihood quadratically and can be marginalized analytically \cite{Elsner:2019rql}. We choose this option as it has shown to reduce the correlation lengths of sampling chains \cite{Kostic:2022vok}. The noise amplitudes and the coefficients $\beta_\uop$ are sampled explicitly. We fix the initial conditions $\delta\lin$ to the known ground truth which allows us to perform tests at a high precision while simultaneously scanning through a wide range of analysis configurations, see also section~\ref{sec:results_precision}. We report our results in terms of mode-centered 68\% credible intervals; further details on the sampling algorithm, stopping criteria and convergence tests are provided in appendix~\ref{sec:sampling-convergence}.

\subsection{Results}
\label{sec:results_results}

\begin{figure}
\centering
\includegraphics[]{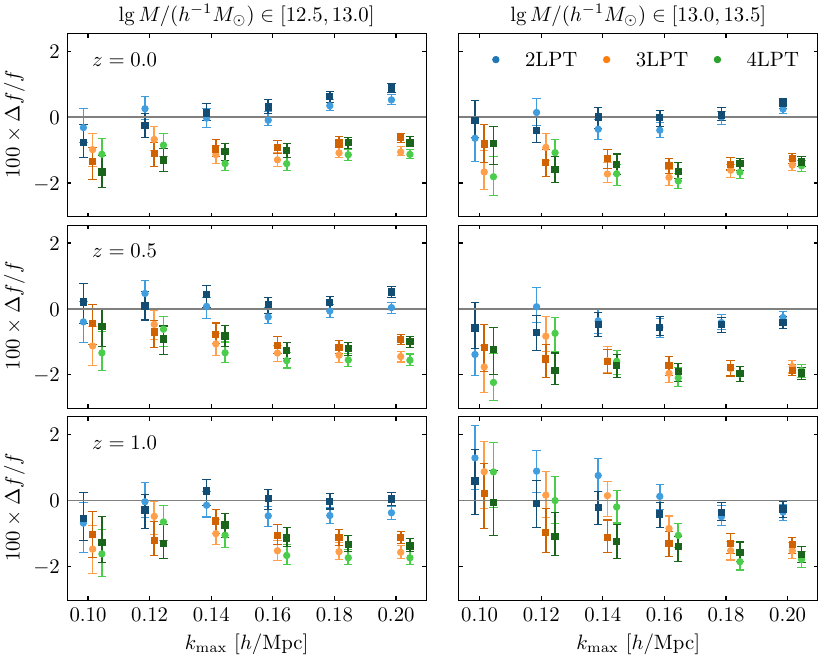}
\caption{Inference of the growth rate $f$ from N-body halos at fixed initial conditions. Constraints are shown in terms of the mode-centered 68\% credible intervals. Two realizations of the simulations available, which are analyzed independently and indicated by different symbols. We compare results at different LPT orders in the gravitational solution for a third order Lagrangian bias expansion with the leading-order velocity bias term and set $\Lambda=k_\mathrm{max}$.}
\label{fig:inference__onestep-varLPT-allz}
\end{figure}

\begin{figure}
\centering
\includegraphics[]{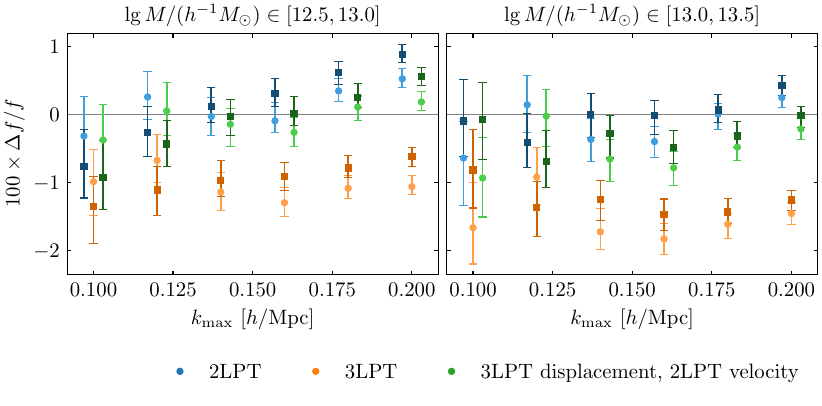}
\caption{Inference of the growth rate $f$ from N-body halos at $z=0$, similar to the first row of figure~\ref{fig:inference__onestep-varLPT-allz}. We compare the 2LPT and 3LPT gravity models with a version that obtains the Lagrangian displacement at third order, but the velocity only at second. Parameter discrepancies in the 3LPT model seem largely driven by the velocity.}
\label{fig:inference__onestep-nlpt-velocityimpact}
\end{figure}

The baseline results for the growth rate inference at fixed initial conditions are shown in figure~\ref{fig:inference__onestep-varLPT-allz}, where we explore both halo mass bins and all three redshift slices for a third order Lagrangian bias expansion and the leading-order velocity bias term (see appendix~\ref{sec:bias-operators} for details). Further, we compare how different LPT orders impact the result. Overall, we obtain constraints that deviate from the ground truth by at most $\sim 1.5\%$, while the standard deviation ranges from 1\% to 0.2\%, depending on the halo sample and cut-off scale. Between the 2LPT and 3LPT model, the inferred mean values shift by up to $\sim 3\%$, while we observe convergence between the 3LPT and the 4LPT result. Somewhat surprisingly, however, the results at 2LPT are closest to the ground truth. This is in contrast to similar parameter recovery tests in the rest frame \cite{Stadler:2023hea}, where an increased LPT order reduces the systematic bias. To test what drives the parameter shift between the 2LPT and the 3LPT results, we compare them to a model in which the Lagrangian displacement is computed at third order, but the velocities at second order, see figure~\ref{fig:inference__onestep-nlpt-velocityimpact}. The comparison indicates that the larger parameter discrepancies of the 3LPT model are mostly driven by the velocities. We will return to this point at the end of this section.
  
\begin{figure}
\centering
\includegraphics[]{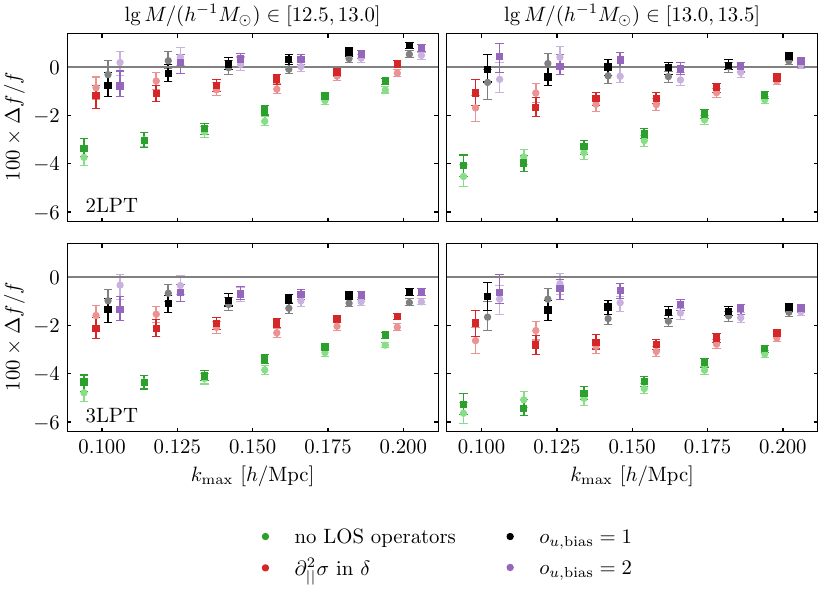}
\caption{Inference of the growth rate $f$ from N-body halos similar to figure~\ref{fig:inference__onestep-varLPT-allz} but now focusing on the 3LPT model at $z=0$. We compare different options for LOS dependent bias terms, namely neglecting all of them, a single $(k\mu)^2$ term in the redshift-space density and velocity operators at leading (one operator) and next-to-leading (four operators) order. Black points in the top/bottom row of this figure (leading-order velocity bias) are identical to the blue/orange measurements from figure \ref{fig:inference__onestep-varLPT-allz}. The lowest-order options shift the inferred value away from the ground truth, while the leading and next-to-leading order velocity bias yield equivalent results. This indicates that LOS-dependent bias terms are important for accurately constraining the growth rate but sufficiently captured by the leading-order velocity bias term.}
\label{fig:inference__onestep-velocitybias}
\end{figure}

\begin{figure}
\centering
\includegraphics[]{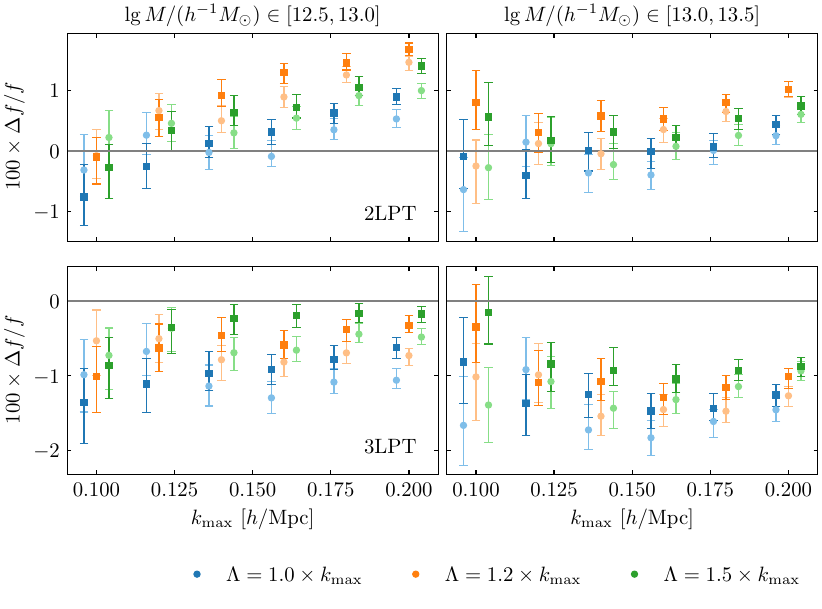}
\caption{Inference of the growth rate $f$ from N-body halos at $z=0$ assuming fixed initial conditions with third order Lagrangian bias, similar to figure~\ref{fig:inference__onestep-varLPT-allz}. We explore the effect of increasing $\Lambda > k_\mathrm{max}$ for the 2LPT and the 3LPT model. The dependency on $\Lambda$ decreases with increasing LPT order, but there is no clear convergence in the inferred value of $f$, it moves further from the ground truth for the 2LPT model and closer in case of the 3LPT model.}
\label{fig:inference__onestep-highL}
\end{figure}

In figure~\ref{fig:inference__onestep-velocitybias}, we investigate how different options to account for velocity bias impact the results. When LOS-dependent operators are excluded entirely from the analysis, the inferred values of $f$ shift significantly away from the ground truth by up to 6\%. One can also expand the redshift-space density in the velocity bias, which at lowest order yields a LOS-dependent bias operator of the form $\left(k\mu\right)^2 \delta$ in the redshift-space galaxy density, see section~\ref{sec:bias-operators-velocity}. The inclusion of only this term yields inferior results compared to the full incorporation of velocity bias at leading order. Increasing the velocity bias to next-to-leading order (i.e. adding three more operators), on the other hand, has no effect. This highlights the importance of higher-derivative bias terms in the velocities, but also indicates that the effect is sufficiently captured by the leading-order velocity operator. The restriction to the leading-order velocity bias reduces the computational demands of the inference at fixed initial conditions significantly. 

To probe the back-reaction from scales beyond $k_\mathrm{max}$ explicitly, we repeat the inference for $\Lambda > k_\mathrm{max}$, see figure~\ref{fig:inference__onestep-higherscaleseffects}. Indeed, we observe a significant effect, in case of the 2LPT model the results move away from the ground truth but for the 3LPT model the consistency improves. In any case, the parameter shifts with increasing $k_\mathrm{max}$ indicate some residual impact from small scales that is not entirely absorbed by the higher-derivative operators (in density and velocity) which are included in the analysis.

We present a set of additional tests in appendix~\ref{sec:additional-tests-inference} and here summarize the results.
\begin{itemize}
\item We investigate the effect of the expansion order $o_\mathrm{bias}$ for the 3LPT model in figure~\ref{fig:inference__onestep-3lpt-varbiasorder}. There is a slight effect in the second mass bin, where the inferred values lie further from the ground truth for $o_\mathrm{bias}=2$. Results at third and fourth order are very consistent among each other and infer $f$ closer to the ground truth. This is the convergence behavior expected for an EFT analysis, but the effect is much weaker than for the $\sigma_8$ inference in the rest frame \cite{Stadler:2024}. Since $\sigma_8$ is perfectly degenerate with linear bias, its heightened dependency on higher-order bias terms is to be expected.
\item We explore two higher-derivative extensions of the bias expansion in figure~\ref{fig:inference__onestep-higherscaleseffects}. Firstly including the operator $\partial^2_\pp u_\pp$ in the velocity expansion. The leading-order impact of such a velocity term on the redshift-space density is proportional to $\mu^4 k^2 \delta$, and a counter-term of this form was introduced in works on the redshift-space power spectrum \cite{Perko:2016puo}. In fact, this term introduces an explicit LOS-dependency in Lagrangian space where the velocity bias expansion is constructed. Further, we allow for higher-derivative operators in the density, that are constructed by application of the Laplace operator to second-order density operators. Neither of these additions has a significant impact on the inferred growth rate.
\item The forward-modeling approach allows us to apply different cuts parallel and perpendicular to the LOS direction and we explore the effect of an anisotropic likelihood filter in figure~\ref{fig:inference__onestep-anisotropicfilter}. Such a filter can exclude the least accurate high wavenumbers closely aligned with the line of sight, and it would also allow to mitigate the Fingers-of-God effect from the random motion of subhalos inside viralized structures. Indeed we observe some improvement for the 3LPT model, and with the most aggressive filter $\left(k\mu\right)^2 \leq \left(0.08\,h/\mpc\right)^2$ the inference becomes consistent with the ground truth, albeit at the price of increasing the error by up to a factor $2$.
\item The impact of using $\sigma = \sum_n \sigma^{(n)}$ in the bias expansion rather than $\sigma^{(1)}$ (see section~\ref{sec:forward-model__summary}) is explored in figure~\ref{fig:inference__sigma-crosschekc}. The parameter shift is generally within the statistical error, and most significant at the highest cutoff values, as expected given that both bias formulations differ only via higher-order terms.
\item We explicitly explore the numerical convergence of our results in figure~\ref{fig:inference__onestep-ngeul}, by increasing $N_\eul$ from $3/2\, N_{\grid,\nyquist}\left(\Lambda\right)$ to $2\, N_{\grid,\nyquist}\left(\Lambda\right)$. As with the rest-frame results \cite{Stadler:2024}, we note no impact and conclude that our choice of $N_\eul$ provides sufficient numerical accuracy while optimizing the computational requirements.
\item All results presented in this section so far neglect the transverse contribution in the 3LPT and the 4LPT forward model, which was found to have a negligible impact on the model accuracy in section~\ref{sec:forward-model__accuracy}. We verify that this also holds for the growth-rate inference in figure~\ref{fig:inference__onestep-curl}.
\end{itemize}

When comparing to the galaxy rest-frame analysis \cite{Stadler:2024}, the analysis in redshift space shows a larger dependence on the LPT order. Since we noted a somewhat slower convergence for the momentum and velocity calculation in section~\ref{sec:velocity}, it appears plausible that a higher LPT order is required for an analysis in redshift space. Indeed, the fact that the power spectrum of residuals in figure~\ref{fig:redshiftspace-accuracy__lpterror_z050} monotonically decreases with increasing LPT order would imply that one expects more accurate results for higher $n_\lpt$. The finding that the 2LPT model yields a smaller systematic bias in the growth-rate inference than higher LPT orders is thus surprising. Given the validation results on the velocity and momentum predictions of 2LPT and 3LPT, we believe that the improved consistency of the 2LPT results is most likely a coincidence. It might be driven by the fact that the perturbative error of the 2LPT power spectrum, albeit larger, shows a weaker $\mu$ dependence, cf. figure~\ref{fig:redshiftspace-accuracy__lpterror-restframecomparison_z050}. In fact, the error even reduces for the highest $k$ modes closely aligned with the line of sight, which we expect to have the largest effect on the inference of $f$.

Further, the dependence of the inferred growth rate on $\Lambda$ at fixed $k_\mathrm{max}$ indicates some residual impact of small-scale modes beyond $k_\mathrm{max}$. This effect is not absorbed by the higher-order bias terms which we explore in figures~\ref{fig:inference__onestep-velocitybias}, \ref{fig:inference__onestep-3lpt-varbiasorder} and \ref{fig:inference__onestep-higherscaleseffects}. It might point to an issue with the simplified noise model (likelihood) adopted here. Specifically, we neglect noise contributions from the RSD displacement in eq.~(\ref{eq:forward-model__rsd-noise}). At next-to-leading order, these take the form $(\partial_{x_\pp} u_\pp)(\vecs{x})\,\epsilon(\vecs{x})$; the term directly correlates with the velocity and depends linearly on the parameter of interest, $f$. It is plausible that this noise term, if unaccounted, impacts the result.  We plan to develop a noise model for \leftfield\ that can account for scale dependence and the noise properties in redshift space simultaneously. This will allow us to explore the dependency of the remaining percent-level systematics on the noise description. We also plan to investigate the convergence of the forward model and the parameter inference for higher LPT orders.

\subsection{On the precision of the growth rate inference}
\label{sec:results_precision}

The analysis of simulated data allows us to fix the initial conditions $\delta\lin$ to the known ground truth. Unbiased cosmological constraints at fixed initial conditions are an important prerequisite for the full cosmological analysis, but they do not automatically guarantee consistent results when the initial conditions are varied simultaneously. The simultaneous inference of cosmological parameters and $\delta\lin$ was demonstrated recently in the rest frame \cite{Kostic:2022vok, beyond2pt, Nguyen:2024yth}, and we plan to explore such an analysis in redshift space in the near future.

Fixing the initial conditions removes cosmic variance to the largest possible extent. The high precision allows to test for small parameter shifts with respect to the ground truth, i.e. to assess the accuracy of the model carefully. In spirit, the test is similar to power- and bispectrum validation tests which average the data vector over multiple simulation realizations and rescale the covariance accordingly, e.g. \cite{Nishimichi:2020tvu, Ivanov:2021kcd, 2024A&A...687A.216E, Eggemeier:2025xwi, Maus:2024sbb, Maus:2024dzi, Noriega:2024eyu, Lai:2024bpl}. The EFT data challenge \cite{Nishimichi:2020tvu}, for example, utilizes a $566~\left(\mathrm{Gpc}/h\right)^3$ volume. In contrast to our analysis, the teams fix the cosmology to a $\Lambda$CDM model, such that the growth rate $f$ is determined by $\Omega_\mathrm{m}$ which in turn is measured by shape and distance. In this setup, RSDs mainly constrain the fluctuation amplitude $A_\mathrm{s}$, which is measured with $0.7\%$ precision at $k_\mathrm{max}=0.12\,h/\mpc$, $z=0.61$ in case of the EFT data challenge. Our results are at a comparable precision, as we obtain $\sigma(f)/f \simeq \left(0.6 - 0.7\right)\%$ at $k_\mathrm{max}=0.12\,h/\mpc$ and $z=0.5$, however, with a much smaller volume of only $8\,(\mathrm{Gpc}/h)^3$.

\begin{figure}
\centering
\includegraphics[]{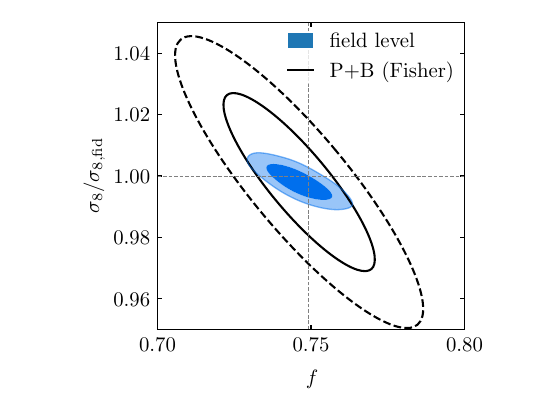}
\caption{Comparison between field-level and power- and bispectrum constraints at fixed initial conditions. We consider the lower halo mass bin at $z=0.5$, $k_\mathrm{max}=0.12\,h/\mpc$. For the field-level constraints, unlike the rest of the paper, we simultaneously infer the linear density-field amplitude $\sigma_8$ and growth rate $f$ while marginalizing over noise amplitudes and the full set of third-order bias parameters. For the power- and bispectrum we perform a Fisher analysis where the covariance has been adjusted to account for fixed initial conditions, and adopt and marginalize over a bias expansion up to second order and only the leading noise term. We include the power spectrum monopole, quadrupole and hexadecapole as well as the bispectrum monopole in the data vector. In both cases, we use the 3LPT forward model with leading-order velocity bias and set $\Lambda=k_\mathrm{max}$.}
\label{fig:results_precision}
\end{figure}

What our tests do not yet answer is the constraining power of field level analysis when applied under realistic conditions, that is inferring initial conditions and cosmology simultaneously from volumes that correspond to current stage IV surveys. So far, the information accessible to field-level analysis versus state-of-the-art power- and bispectrum (P+B) measurements has, in the context of the EFT approach, only been investigated in the restframe \cite{Cabass:2023nyo, Nguyen:2024yth, Schmidt:2025iwa, Spezzati:2025zsb} (see \cite{ramanah:2019} for related field-level results in redshift space using a heuristic galaxy forward model). While we leave a full investigation of the redshift-space field-level information for future work, we can estimate the constraining power of the P+B analysis for our setup by performing a Fisher forecast. In the forecast, we use a modified covariance that accounts for fixed initial conditions \cite{Babic:2022dws}. We further restrict the bias expansion to second order for the P+B analysis, see section~\ref{sec:bias-operators-density}, including the next-to-leading order derivative operator $\nabla_q^2\sigma$, the leading velocity operator (eq.~\ref{eq:bias-operators__leading-order-velocity}) and the leading white noise term from eq.~(\ref{eq:forward-model__noise}). Their fiducial values are set to match the lower halo mass bin at $z=0.5$ and obtained by a field-level measurement at fixed cosmology. For the field-level results we consider the full third-order bias expansion with scale- and direction-dependent noise terms.\footnote{Specifically, we consider the 3LPT results of figure~\ref{fig:inference__onestep-varLPT-allz} in this comparison.} Focusing on the lower mass bin at $k_\mathrm{max}=0.12\,h/\mpc$, $z=0.5$, we obtain, at fixed $\sigma_8$, $\sigma_\mathrm{P+B}(f)/f_\mathrm{fid} = 1.1\%$ in comparison to the $0.6\%$ measurement by the field-level analysis. 

Notice that up to this point and throughout this paper, we have kept the amplitude of initial fluctuations $\sigma_8$ fixed, as our goal was to perform a rigorous test of the redshift-space modeling. The joint inference of $\sigma_8$ and $f$, corresponding in a sense to a consistency test of GR, yields even more substantial gains from the field-level analysis over power spectrum and bispectrum (figure~\ref{fig:results_precision}). This gain is obtained due to the effective breaking of the linear-order degeneracy between the two parameters present for biased tracers. We emphasize again that these results, while a fair comparison, are strictly for the case of fixed initial conditions.

\section{Conclusions}
\label{sec:conclusions}

This work introduces a revised model for redshift space distortions in \leftfield. Its most important improvements are better numerical convergence and faster runtime. This is achieved by going from Lagrangian frame to redshift space in a single displacement step whereby the shift vectors from the gravitational displacement and the peculiar velocity are added. By treating higher-order contributions to the shift vector non-perturbatively we keep symmetry-protected shift terms to arbitrarily high order. In the future, the approach will allow to directly predict the galaxy density on the light cone, and to incorporate the distance rescaling to a fiducial cosmology similar to \cite{ramanah:2019}.

We evaluate the forward model against reference N-body simulations with an identical Fourier-space top-hat filter at the scale $\Lambda$ applied to their initial conditions and find good agreement. The power spectrum can be predicted to better than 2\% accuracy at $\Lambda=0.20\,h/\mpc$ and $z=0.5$, and it further improves to sub-percent accuracy at $\Lambda=0.10\,h/\mpc$. Overall, the redshift-space forward model is found to be slightly less accurate than its rest-frame companion \cite{Stadler:2024}. This can be explained by the somewhat slower perturbative convergence of momentum and velocity in Lagrangian Perturbation Theory, which essentially are time derivatives of the density.

We perform a series of parameter recovery tests at the field level from simulated halo data sets which contain the full nonlinearity and non-trivial bias properties expected for realistic tracers. To increase the precision of those tests and to be able to scan through a wide range of analysis configurations, we fix the initial conditions to the known ground truth. The growth rate is inferred with $0.2 - 1.0\,\%$ statistical precision, allowing us to identify a remaining systematic parameter shift at the percent level. This systematic, which shrinks when increasing the cutoff $\Lambda$ relative to $k_{\rm max}$, is linked either to the velocity prediction, or the noise model adopted. We plan to investigate the issue further in the future and to develop a noise model which can simultaneously account for scale dependence and anisotropic noise properties in redshift space.

The speed-up achieved for the redshift-space modeling now enables further progress in two directions. First, the model is fast enough to be employed in field-level analyses of the redshift-space galaxy density. The higher-order information contained in the density allows to break linear-order degeneracies more efficiently than the combination of power- and bispectrum \cite{beyond2pt,Nguyen:2024yth}. This is illustrated for the case of fixed initial conditions in figure~\ref{fig:results_precision}. We plan to explore the full potential of field-level analyses to put independent constraints on $f$ and $\sigma_8$ while marginalizing over bias and initial conditions in the future. Second, given the speedup, simulation-based inference in redshift space becomes possible. The planned implementation of a survey mask will elevate the forward model to the complexity of current EFT-based power- and bispectrum analyses and facilitate SBI constraints from realistic light cones.

\acknowledgments{We thank Adam Andrews for insightful feedback on this manuscript and Ivana Babi\'c, Safak Celik, Ivana Nikolac, Minh Nguyen, Moritz Singhartinger and Beatriz Tucci for useful discussions. This work  was funded by the Deutsche Forschungsgemeinschaft (DFG, German Research Foundation) under Germany's Excellence Strategy – EXC 2094 – 390783311”.}

\appendix

\section{Resolution effects in the Eulerian-frame momentum}
\label{sec:resolution-effects__momentum}

We expect resolution effects in the Eulerian-frame momentum to behave similarly as the Lagrangian bias operators that were investigated in ref.~\cite{Stadler:2024}; both are obtained from the displacement of a weight field. The most decisive parameter for their numerical accuracy is $N_\eul$, the particle number and grid size in the displacement operation. For the simplest case $k_\nyquist\left(N_\eul\right) = \Lambda$, which is the minimum grid size to represent all modes up to the cut-off $\Lambda$, residuals due to resolution effects are expected to scale as (see section 3.5 of ref.~\cite{Stadler:2024})
\begin{equation}
\left\langle \left(\Delta \pi_\pp\right)^2\right\rangle^\LO \propto k^4 \,.
\end{equation} 
With increasing resolution the residuals decrease rapidly, and $N_\eul = 3/2\, N_{\grid,\nyquist}\left(\Lambda\right)$ has proven suitable to minimize resolution effects at acceptable computational speed.

\begin{figure}
\centering
\includegraphics[]{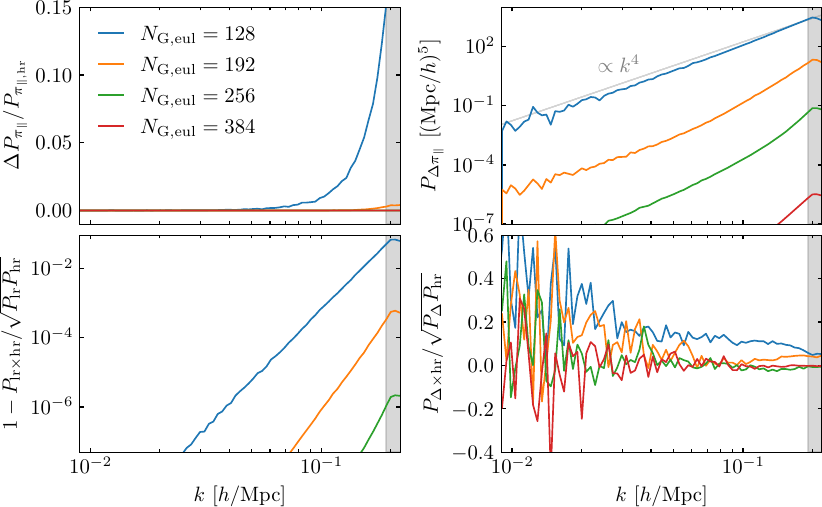}
\caption{Resolution effects in the Eulerian-frame momentum  $\pi_\pp\left(\vec{x}\right)$. The forward model computes the displacement at third order in the LPT solution from initial conditions filtered at $\Lambda=0.20\,h/\mpc$ and neglects the transverse contribution. We compare results at different displacement resolutions (subscript ``lr'') to a reference model that has $N_\eul=768$ (subscript ``hr''); the comparison is shown in terms of the power spectrum difference, the residuals power spectrum, the cross-correlation between residuals and reference and the cross correlation between low-resolution model and reference (clockwise from the top left). The results are extended to different LPT orders in figure~\ref{fig:velocity-accuracy__resolution-effects-momentum-lptorder}.}
\label{fig:velocity-accuracy__resolution-effects-momentum}
\end{figure}

\begin{figure}
\centering
\includegraphics[]{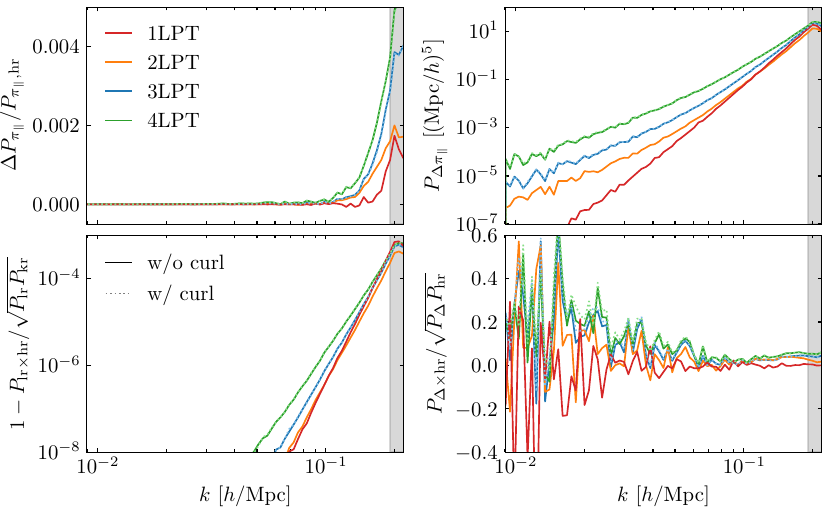}
\caption{Resolution effects in the Eulerian-frame momentum density $\pi_\pp\left(\vec{x}\right)$ similar to figure~\ref{fig:velocity-accuracy__resolution-effects-momentum}. We compare results at a cut-off $\Lambda=0.20\,h/\mpc$ and nominal resolution $N_\eul=3/2\, N_{\grid,\nyquist}\left(\Lambda\right) = 192$ to a high-resolution reference with $N_\eul=768$ for different LPT orders.}
\label{fig:velocity-accuracy__resolution-effects-momentum-lptorder}
\end{figure}

We test numerical effects in $\pi_\pp$ in figure~\ref{fig:velocity-accuracy__resolution-effects-momentum} by comparing the forward model prediction at different resolutions $N_\eul$ to a high-resolution reference. Indeed, we recover the expected $k^4$ scaling and see a rapid decrease of the residuals with increasing resolution. Also the cross-correlation between residuals and ground truth decays quickly. In figure~\ref{fig:velocity-accuracy__resolution-effects-momentum-lptorder} we extend the comparison to different LPT orders, and we also consider the transverse contribution to the displacement. There is a small impact of the LPT order on the numerical accuracy, with higher orders implying slightly larger residuals because they populate higher wavenumbers more strongly. The transverse displacement contribution, on the other hand, has no recognizable impact on the numerical accuracy. Overall, we find that resolution effects in the momentum density behave as expected and can be well controlled by the setting $N_\eul = 3/2\, N_{\grid,\nyquist}\left(\Lambda\right)$, i.e. the same rule that applies to the density field and Lagrangian bias operators.

\section{Velocity estimation by the Delaunay tessellation}
\label{sec:delaunay}

Estimation of the velocity field from N-body simulations is notoriously difficult because the simulation particles trace the density, and in low density regions the velocity estimations suffers large Poisson errors or even becomes undetermined (see e.g. refs.~\cite{Juszkiewicz:1993hm, Bernardeau:1995en, Bernardeau:1996hb}). For this reason, different techniques have been proposed to reconstruct the velocity field in the voids between particles (see e.g. refs.~\cite{Hahn:2014lca, Bel:2018awq, Feldbrugge:2024wcm, Esposito:2024qlo} for some recent applications). A common choice to estimate the volume-weighted velocity is by a Delaunay tessellation field estimator; here we employ the method from ref.~\cite{Bel:2018awq}, which is closely adapted from ref.~\cite{Pueblas:2008uv}. The Delaunay tessellation divides the simulation box into tetrahedra with a particle at each vertex such that there is no other particle in the circumsphere of any  tetrahedron, i.e. the tetrahedra are optimized to be as equiangular as possible. The velocity field is interpolated over the entire space by assuming it to be continuous with a constant gradient over each tetrahedron. The velocity assigned at each grid node then is given as the volume-weighted average velocity over a spherical cell with the node at its center,
\begin{equation}
\vec{v}\left(\vec{x}_i\right) = \frac{\sum_{t=1}^{N_t} w_t \vec{v}_t}{\sum_{t=1}^{N_t} w_t} \,,
\end{equation}
where $N_t$ is the number of tetrahedra with at least one vertex inside the sphere, $w_t$ the volume of the tetrahedron inside the sphere and $\vec{v}_t$ the average of the linearly interpolated velocity field over the tetrahedron inside the sphere. For tetrahedra entirely contained in the cell, $\vec{v}_t$ is the arithmetic mean of the four velocities at the vertices For tetrahedra that intersect the cell only partly, $w_t$ and $\vec{v}_t$ are estimated from a set of random points drawn uniformly over the tetrahedron. Their number is proportional to the volume of the tetrahedron an bound from below by the parameter $N_\mathrm{rand.}$. The size of the spherical cell and hence the smoothing of the velocity field is set by the size of the Delaunay grid, $R_\mathrm{cell} = L_\mathrm{box}/N_\delaunay$. In practice, we perform the Delaunay tessellation locally around each cell to cope with the large number of particles in the simulations. We choose this local region to ensure that the Delaunay tetrahedra cover at least the fraction $f_\mathrm{cell}$ of each spherical cell \cite{Bel:2018awq}.

In section~\ref{sec:velocity_velocity-accuray} we apply the Delaunay tessellation algorithm to particles generated from the \leftfield\ forward model and to the reference simulations with the following parameters
\begin{equation}
N_\delaunay = 256\,, \quad
N_\mathrm{rand.} = 25\,, \quad
f_\mathrm{cell} = 93\% \,.
\end{equation}
The algorithm accepts input files in the format of \texttt{GADGET} snapshots \cite{Springel:2005mi, Springel:2020plp} and can be applied directly to the reference simulations. For \leftfield, we generate an ensemble of uniformly distributed particles which are weighted by the Lagrangian velocity (eq.~\ref{eq:velocity_velocity}) at the respective positions, then we move the particles by the LPT displacement vector (eq.~\ref{eq:velocity_eulerian-lagrangian}) to the Eulerian frame. Rather than assigning the particles to a rectangular grid, which would be the next step in the forward model, we write the positions and velocities to disk and use them as input to the Delaunay code.

\begin{figure}
\centering
\includegraphics[]{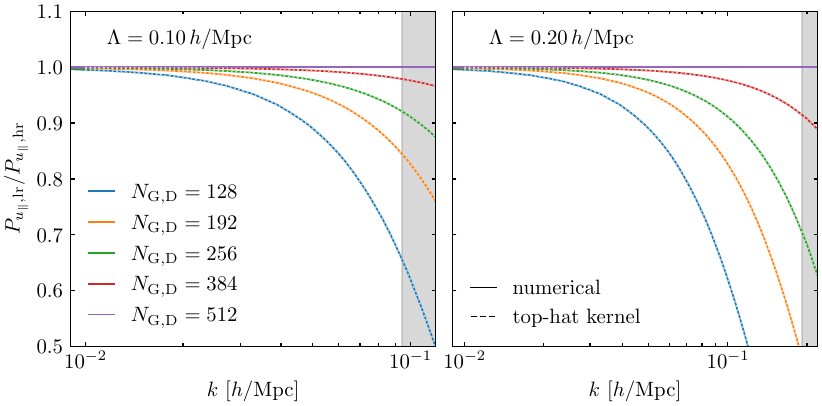}
\caption{Effect of the target grid size $N_\delaunay$ on the velocity power spectrum estimated by the Delaunay tessellation. The Delaunay tessellation is applied to simulations with a cut-off in the initial conditions at $\Lambda = 0.10,\,0.20\, h/\mpc$, respectively and $N_\mathrm{part.}=1536^3$ particles. We compare the resulting velocity power spectrum at target resolutions $N_\delaunay$ (subscript ``lr'') to a high-resolution reference with $N_\delaunay=512$ (subscript ``hr''). Dashed lines indicate expected smoothing by a top-hat kernel at the respective resolutions, and they well describe the smoothing effect of $N_\delaunay$.}
\label{fig:velocity-accuracy__delaunay-grid-size}
\end{figure}

\begin{figure}
\centering
\includegraphics[]{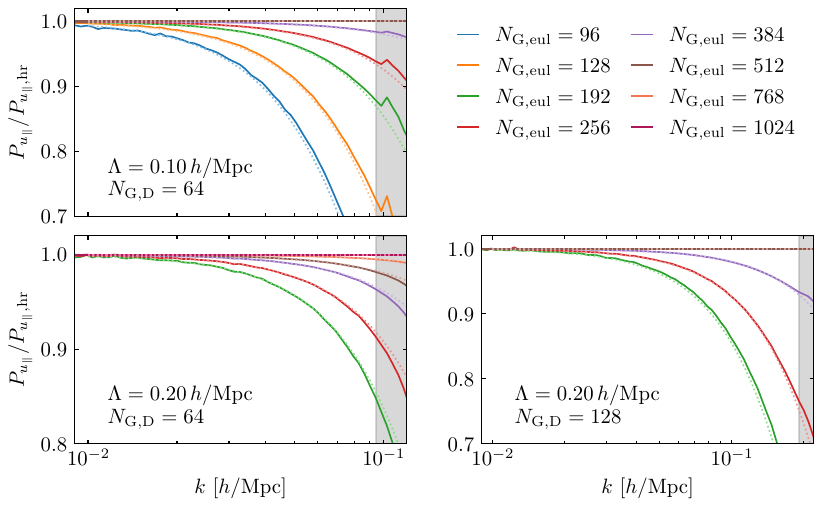}
\caption{Impact of the particle number on the velocity power spectrum estimated by the Delaunay tessellation. We use the \leftfield\ forward model and sample the velocity field with different numbers of particles $N_\mathrm{part.}=N_\eul$, then assign to the same Delaunay grid size $N_\delaunay$ and compare the results to a high resolution reference with $N_\eul=1024$ (subscript ``hr''). Dashed line indicate the smoothing by a CIC kernel at the respective resolutions. Note that we always choose $N_\eul \geq 3/2\, N_{\grid,\nyquist}\left(\Lambda\right)$ to suppress resolution effects in the perturbative calculation.}
\label{fig:velocity-accuracy__delaunay-particle-number}
\end{figure}

Both the target grid size of the Delaunay tessellation, $N_\delaunay$, and the particle number in the simulation $N_\mathrm{part.}$ imply a smoothing of the output velocity field. Given the algorithm, we expect the grid size to act as a top-hat filter of radius $R_\mathrm{cell}$, and indeed we recover this behavior in figure~\ref{fig:velocity-accuracy__delaunay-grid-size}. There, we keep $N_\mathrm{part.}$ constant and compare the velocity power spectrum for target grids of different resolutions. In figure~\ref{fig:velocity-accuracy__delaunay-particle-number}, we investigate the impact of $N_\mathrm{part.}$, whereby the \leftfield\ model allows us to quickly generate non-linear velocity fields at different resolutions. We find that the smoothing implied by the particle number is well approximated by a CIC kernel of width $L_\mathrm{box}/N_\mathrm{part.}$. 

When comparing the \leftfield\ model and the reference simulations in section~\ref{sec:velocity_velocity-accuray}, we choose equal $N_\delaunay=256$ such that the smoothing due to the grid sizes cancels. The particle number of simulations and forward model are $1536^3$ and $1024^3$, respectively, where we have chosen the latter for the sake of computational efficiency. The expected impact on the power spectrum from differing particle numbers is indicated in figures~\ref{fig:velocity-accuracy__perturbative-accuracy-velocity}, \ref{fig:velocity-accuracy__accuracy-velocity-momentum-density_L010_z050}, and~\ref{fig:velocity-accuracy__accuracy-velocity-momentum-density_z100}, and it is smaller than the perturbative accuracy for all scenarios that we consider.

\section{Operators in the bias expansion}
\label{sec:bias-operators}

\subsection{Density operators}
\label{sec:bias-operators-density}

For the bias expansion we consider the following leading-order in derivative operators:
\begin{itemize}
\item First order: $\sigma$
\item Second order: $\sigma^2$, $\tr \left[ M^{(1)}\,M^{(1)}\right]$ \,,
\item Third order: $\sigma^3$, $\sigma\,\tr \left[M^{(1)}\,M^{(1)}\right]$, $\tr \left[M^{(1)}\,M^{(1)}\,M^{(1)}\right]$,  $\tr \left[M^{(1)}\,M^{(2)}\right]$ \,,
\item Fourth order: $\sigma^4$, $\sigma^2\,\tr \left[M^{(1)}\,M^{(1)}\right]$, $\sigma\,\tr \left[M^{(1)}\,M^{(1)}\,M^{(1)}\right]$, $\sigma\,\tr \left[M^{(1)}\,M^{(2)}\right]$,\newline\phantom{Fourth order:} $\tr \left[M^{(1)}\,M^{(1)}\,M^{(2)}\right]$, $\left(\tr \left[ M^{(1)}\,M^{(1)}\right]\right)^2$,  $\tr \left[M^{(1)}\,M^{(3)}\right]$,  $\tr \left[M^{(2)}\,M^{(2)}\right]$\,.
\end{itemize}
where $M^{(n)}$ is the symmetric part of the Lagrangian distortion tensor at $n$-th order and $\sigma=\sum_n\sigma^{(n)}$ the divergence of the displacement vector in eq.~(\ref{eq:velocity_displacement-decomposition}). The use of $\sigma$ instead of $\sigma^{(n)}$ is primarily for a more convenient implementation, and we verify in figure~\ref{fig:inference__sigma-crosschekc} that it has no significant impact on the analysis. In addition, all inferences considered in this work marginalize over the next-to-leading order derivative operator $\nabla_q^2 \sigma$.

\subsection{Velocity operators}
\label{sec:bias-operators-velocity}

The velocity bias expansion, \refeq{forward-model__velocity-bias}
\begin{equation*}
\udgpp\left(\vec{q}\right) = u_\pp\left(\vec{q}\right) + \sum_{\{\uop\}} \beta_\uop\, \uop\left(\vec{q}\right)\,,
\end{equation*}
consists of all vectorial quantities that can be constructed from the Lagrangian distortion tensor by contracting it with derivatives and the line of sight direction. Due to the equivalence principle, the linear order term $u_\pp$ is bias free, i.e. $\beta_{u_\pp}=1$. At leading order in the bias expansion, there is one term which we write as
\begin{equation}
\{ \uop \}_{o_{u,\bias}=1} = \{ \partial_{q\pp} \sigma  \}\,.
\label{eq:bias-operators__leading-order-velocity}
\end{equation}
At next-to-leading order, three more terms appear:
\begin{equation}
  \{ \uop \}_{o_{u,\bias}=2} =
  \left\{
\partial_{q_\pp} \sigma^2\,, \quad
\partial_{q_\pp} \tr\left[\left(M^{(1)}\right)^2\right] \,, \quad
M^{(1)}_{\pp,j}\, \partial_{q_i} M^{(1)\,ij}
\right\}\,.
\label{eq:bias-operators__higher-order-velocity}
\end{equation}
For a more convenient implementation, we replace the last term by
\begin{equation}
\partial_{q_i} \left[M^{(1)}_{\pp,j} M^{(1)\,ij} \right] = \frac{1}{2} \partial_{q_\pp} \tr\left[\left(M^{(1)}\right)^2\right] + M^{(1)}_{\pp,j}\, \partial_{q_i} M^{(1)\,ij} \,,
\end{equation}
which corresponds to a simple redefinition of the bias coefficients. All above derivatives are taken in Lagrangian coordinates, and up to the order considered here Lagrangian and Eulerian derivatives are equivalent. Concretely, expanding eq.~(\ref{eq:forward-model__eulerian-derivatives}) to second order for the leading velocity operator yields
\begin{equation}
\frac{\partial \sigma^{(1)}}{\partial x_\pp} \simeq \frac{\partial \sigma^{(1)}}{\partial q_\pp} - M^{(1)}_{\pp, i} \frac{\partial \sigma^{(1)}}{\partial q_i} \,.
\end{equation} 
That is, the difference between Lagrangian and Eulerian derivatives arises one order beyond the original operator and can be absorbed by the next-to-leading-order velocity bias terms.

Finally, it is instructive to consider how the velocity operators affect the density contrast in redshift space. To this end, the displacement of eq.~(\ref{eq:intro__rsd-displacement}) can be expanded as \cite{Desjacques:2016bnm},
\begin{equation}
\dgds\left(\vecs{x}\right) = \dgd\left(\vec{x}\right) + \sum_{n=1}^{\infty} \frac{(-1)^n}{n!} \partial^n_\pp \left[ \udgpp^n\left(\vec{x}\right) \left( 1 + \dgd\left(\vec{x}\right)\right)\right] \,.
\end{equation}
The leading-order velocity bias operator, $\partial_\pp \sigma^{(1)} \propto \partial_\pp \delta^{(1)}$ enters the redshift-space density contrast as $\partial^2_\pp \delta^{(1)} \propto \left(k\mu\right)^2 \delta^{(1)}$. That is, it correctly yields the leading-order redshift-space counter term \cite{Perko:2016puo}. A term of the form $\mu^4 k^2 \delta^{(1)}$ would arise from a velocity bias operator $\partial_\pp^2 u_\pp$. Formally, this term corresponds to introducing a preferred direction in the tracer rest frame and hence a selection effect in the sense of ref.~\cite{Desjacques:2018pfv}. We explore the impact of such a term on the inference in figure~\ref{fig:inference__onestep-higherscaleseffects} and find it to be negligible.

\section{Comparison with the two-step model}
\label{sec:two-step-comparison}

The redshift-space forward model of this work is summarized in figure~\ref{fig:redshiftspace-model__flowchart}. Importantly, it takes a single displacement step from Lagrangian coordinates to redshift space by adding up the LPT displacement vector and the velocity. We therefore refer to the model as ``one-step model''. In a previous publication, we explored modeling redshift space distortions in a two-step procedure, where we shift the operators from Lagrangian to Eulerian coordinates, obtain the velocity in the Eulerian frame, and perform a second displacement. We refer to this treatment as ``two-step model''. One immediate advantage of the one-step model is its computational efficiency, by saving the second displacement operation.

The different treatment of the displacement to redshift space also has some implications for the bias expansion, because operators in the Eulerian frame can not be directly accessed in the one-step model. In particular, the linear bias in the two-step model is
\begin{equation}
\frac{b_1 \delta_\matter\left(\vecs{x}\right)}{1 + \partial_{x_\pp} \udgpp\left(\vecs{x}\right)} \,,
\end{equation}
while in the one-step model we have the expression given in eq.~(\ref{eq:forward-model__linear-bias}). Further, higher-derivative operators in density and velocity are computed in the Lagrangian frame for the one-step model but in Eulerian coordinates for the two-step model. We argue in section~\ref{sec:forward-model__summary} and appendix~\ref{sec:bias-operators-velocity} that either option should be equivalent up to the operator order considered in this paper. Note also that, in the two-step model, the leading-order velocity operator (eq.~\ref{eq:bias-operators__leading-order-velocity}) is replaced by $\partial_{x_\pp}\delta_\matter$. In the following, we compare the one-step and the two-step model in terms of their numerical accuracy (section~\ref{sec:two-step-comparison-modelaccuracy}) and the parameter inference (section~\ref{sec:two-step-comparison-inference}).

\subsection{Numerical accuracy}
\label{sec:two-step-comparison-modelaccuracy}

\begin{figure}
\centering
\includegraphics[]{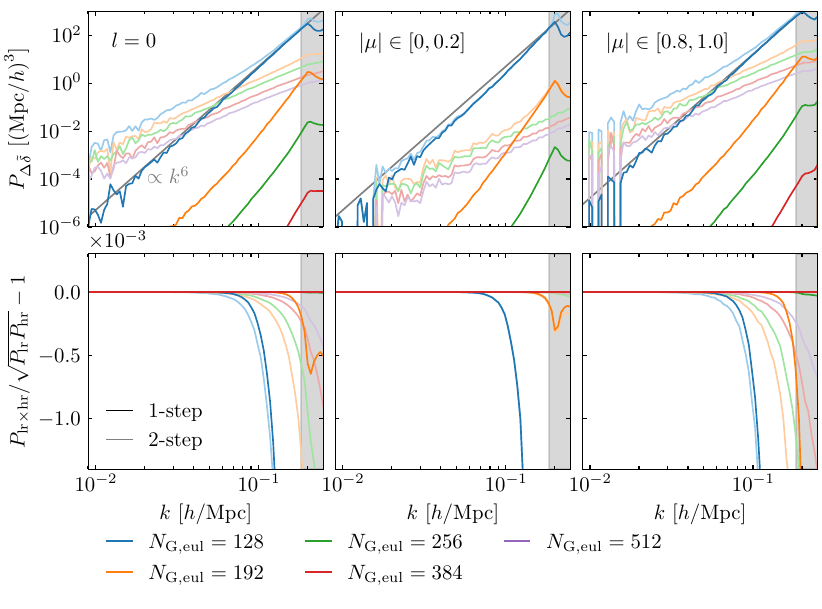}
\caption{Resolution effects in the redshift-space forward model. We consider the one-step (bold) and the 2-step (pastel) 3LPT model for $\Lambda=0.20\,h/\mpc$ and compare results at different resolutions of the displacement step to a high-resolution reference with $N_\eul=1024$. Differences are presented in terms of the power spectrum monopole (left) and two wedges perpendicular (middle) and parallel (right) to the line of sight. For corresponding results at different LPT orders with the one-step model see figure~\ref{fig:redshiftspace-accuracy__resolution-nLPT-summary}. For the one-step model, we recover the expected $k^6$-scaling, and we observe a rapid decay of the residuals as the resolution of the displacement increases. In contrast, the two-step model converges much slower and has a shallower scaling of the residuals.}
\label{fig:redshiftspace-accuracy__resolution-3LPT-summary}
\end{figure}

For the one-step model, the same theoretical considerations about resolution effects hold as in the rest frame \cite{Stadler:2024}, and we expect the residuals to scale $\propto k^6$, see eq.~(\ref{eq:redshiftspace-model__resolution-residuals}). Indeed, we recover this expected behavior in figure~\ref{fig:redshiftspace-accuracy__resolution-3LPT-summary}. In comparison, the two-step model exhibits a much shallower scaling of the residuals. Even more importantly, the residuals of the two-step model decrease much less rapidly as the resolution of the displacement step $N_\eul$ is increased. Even for resolutions which become computationally challenging in the two-step model ($N_\eul \gtrsim 384$), the residuals exceed the one-step model with $N_\eul = 3/2\, N_{\grid,\nyquist}\left(\Lambda\right)$. This is the case in particular for modes close to the LOS direction, where the velocity has the highest impact. Thus from the viewpoint of numerical speed and accuracy, the one-step model is clearly preferred.

\subsection{Analysis results}
\label{sec:two-step-comparison-inference}

In a previous study \cite{Stadler:2023hea}, we explored the two-step model and obtained percent-level constraints on the growth rate $f$ at fixed initial conditions. As data set, we considered halos identified from the same N-body simulations as used in this work, but with the AMIGA \cite{2009ApJS..182..608K} halo finder rather than with ROCKSTAR \cite{Behroozi2013}. Another important difference is that the density assignment in the forward model was performed with a CIC kernel, rather than with the NUFFT algorithm employed here. To facilitate a direct comparison between the one-step and the two-step model, we here repeat the two-step analysis for the ROCKSTAR data set of this work using the NUFFT assignment scheme. We keep the grid size $N_\eul = 384$ from ref.~\cite{Stadler:2023hea}. Indeed tests with the two-step model and the NUFFT algorithm show that results at smaller resolutions have not converged well, indicating that the large grid size necessary in the two-step model is caused by dividing the Jacobian from the momentum to obtain the velocity rather than by the smoothing from the CIC assignment. 

\begin{figure}
\centering
\includegraphics[]{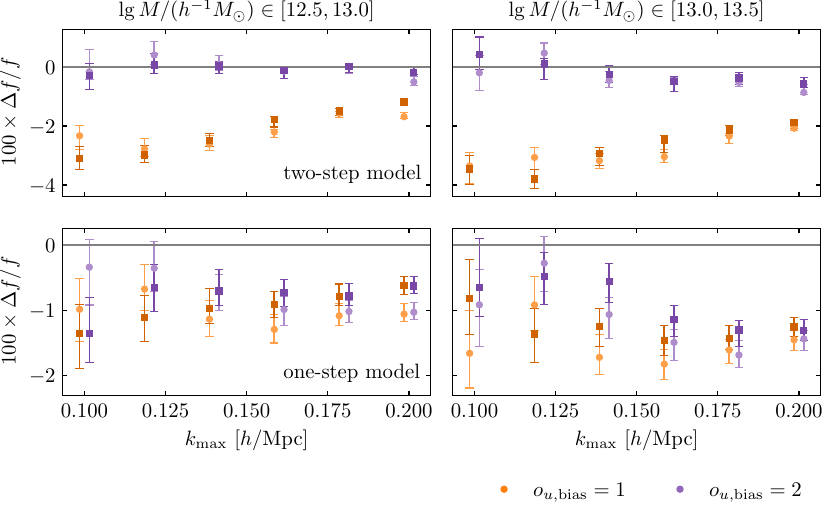}
\caption{Inference of the growth rate $f$ from N-body halos at $z=0.0$ assuming fixed initial conditions. There are two realizations of the simulations available, which are analyzed independently and indicated by different symbols. We use the 3LPT gravity model, third-order Lagrangian bias and set $\Lambda=k_\mathrm{max}$. This is the same configuration as the baseline analysis of ref.~\cite{Stadler:2023hea}, which the exception that we use the NUFFT density assignment rather than a CIC kernel. In the top panel, we investigate the impact of the higher-order velocity operators in eq.~(\ref{eq:bias-operators__higher-order-velocity}) on the two-step model; the bottom panel repeats the corresponding results from figures~\ref{fig:inference__onestep-varLPT-allz} and~\ref{fig:inference__onestep-velocitybias} for the one-step model. While results by the one-step model remain unaffected by higher-order velocity operators, they have a significant impact for the two-step model and make its results more consistent with the ground truth.}
\label{fig:inference__twostep-vs-onestep-3lppt}
\end{figure}

The analysis results of the two-step model and the one-step model are compared in figure~\ref{fig:inference__twostep-vs-onestep-3lppt} for the 3LPT model with next-to-leading-order velocity bias, i.e. the baseline configuration of ref.~\cite{Stadler:2023hea}. We observe that these results are very consistent with the ground truth up to the highest cut-offs tested. Higher-order velocity bias operators have a considerable impact on the two-step model; without their inclusion the result shifts away from the ground truth significantly. The same is not the case for the one-step model, as we have seen in section~\ref{sec:results} and is shown for comparison in figure~\ref{fig:inference__twostep-vs-onestep-3lppt}. This indicates that higher-order velocity operators are able to at least partly absorb the numerical inaccuracies of the two-step model.

\begin{figure}
\centering
\includegraphics[]{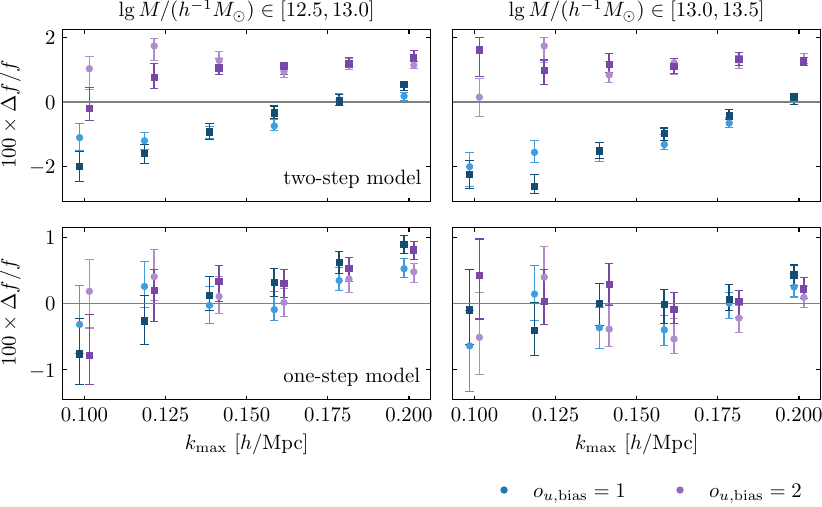}
\caption{Same as figure~\ref{fig:inference__twostep-vs-onestep-2lppt} but now using the 2LPT model for gravity. As previously, we observe that velocity operators beyond the leading order have a significant impact for the two-step model. In contrast, there impact on the one-step model is only very minor.}
\label{fig:inference__twostep-vs-onestep-2lppt}
\end{figure}

We repeat the comparison between one-step and two-step model for a 2LPT description of gravity in figure~\ref{fig:inference__twostep-vs-onestep-2lppt}. We find the same trend as in the one-step case: all other variables being equal, the 2LPT-based model yields a systematically higher inferred growth rate than the 3LPT-based model. Again, next-to-leading order velocity bias is only important in the case of the two-step model. We thus surmise that the same physical velocity-related effect is responsible for the difference between 2LPT and 3LPT in both RSD formulations. 

Finally, an important difference between the one-step and the two-step model are their computational speed. Here, the one-step model profits in three ways. Firstly, each forward model evaluation requires less displacement operations (approximately half as many), which are particularly costly. Second, the improved numerical convergence of the one-step model permits the use of lower displacement resolutions $N_\eul$. And third, the reduction from four to one velocity operators accelerates the inference because the forward model, or more concretely the second displacement step in the two-step model, needs to be re-evaluated for each new value of the velocity coefficients $\beta_\uop$. The former two effects are the more decisive ones for a field-level analysis. With the analysis configurations chosen for figure~\ref{fig:inference__twostep-vs-onestep-3lppt} at $o_{u,\bias}=1$, one evaluation of the one-step model takes about 2\% to 10\% of the time required for the two-step model. Because the resolution $N_\eul$ depends on the cut-off $\Lambda$ in the one-step but not in the two-step settings, the highest speed up is achieved for small $\Lambda$.

\section{Sampling algorithm and convergence}
\label{sec:sampling-convergence}

The (density) bias coefficients $b_\op$ enter the likelihood in eq.~(\ref{eq:forward-model__likelihood}) quadratically and hence can be marginalized analytically \cite{Elsner:2019rql}. This leaves as free parameters in the analysis the growth rate $f$, the noise amplitudes in eq.~(\ref{eq:forward-model__noise}) and the coefficients of the velocity bias operators in eq.~(\ref{eq:bias-operators__leading-order-velocity}) and eq.~(\ref{eq:bias-operators__higher-order-velocity}). Which of the latter precisely are present depends on the analysis choice. We assume flat and uninformative priors for all free parameters, which we list in table~\ref{tab:sampling-priors}. The posterior is explored with the slice sampler \cite{2000physics...9028N}, adopting the ``stepping out'' and ``shrinkage'' procedure, and in a block sampling approach we sample one parameter at a time conditional on all others. This algorithm was implemented directly in \leftfield.

For the one-step model, we run each chain for 1,990 to 2,000 accepted samples, where the exact number slightly differs because we have to remove some overlap after a chain is interrupted and restarted. After inspection of the trace plots, we remove 100 samples for burn in. We demand an effective samples size\footnote{The effective sample size and the Gelman-Rubin criterion are computed with the \texttt{NumPyro} package, \url{https://num.pyro.ai/}} of at least 75 in the noise amplitudes and at least 100 in all other parameters. The latter criterion is easily fulfilled for chains with $\sim 2000$ samples, and we typically obtain an effective sample size between several hundred to more than a thousand for the growth rate. The noise amplitudes exhibit longer correlation lengths, and in the two instances where the criterion of at least 75 effective samples is not met, we continue the chains for another 2,000 accepted samples. 

In case of the two-step model, collecting $\sim 2000$ samples per inference is not feasible due to the higher computational costs. After inspecting the trace plots, we discard the first 50 samples for burn in, and we monitor the effective sample size in $f$. We run the chains until we have collected at least 100 independent samples of $f$.

\begin{table}[h]
\centering
\begin{tabular}{c|c|c|c|c|c}
parameter & lower limit & upper limit & step size & initial value & initial range \\
\hline
$f$                          & $0.1$      & $2.0$     & $0.01$ & $0.513$ & $0.3 - 1.0$\\
$\sigma_0$                   & $0.0$      & $10^{30}$ & $0.01$ & $0.3$   & $0.1 - 2.0$ \\
$\sigma_{\epsilon,2}$        & $0.0$      & $10^{30}$ & $3.0$  & $0.0$   & $0 - 50$\\
$\sigma_{\epsilon\mu,2}$     & $0.0$      & $10^{30}$ & $4.0$  & $0.0$   & $0 - 50$\\
$\beta_{\partial_\pp\sigma}$          & $-10^{30}$ & $10^{30}$ & $8.0$ & $0.0$   & $-50 - 50$\\
$\beta_{\partial_\pp\sigma^2}$         & $-10^{30}$ & $10^{30}$ & $12.0$ & $0.0$   & --\\
$\beta_{\partial_\pp\tr[(M^{(1)})^2]}$ & $-10^{30}$ & $10^{30}$ & $25.0$ & $0.0$   & --\\
$\beta_{\partial_\pp M^{(1)}_{\pp,j} \partial_i M^{(1),ij} }$ & $-10^{30}$ & $10^{30}$ & $23.0$ & $0.0$   & --\\
\end{tabular}
\caption{Free parameters whose posteriors are explored with a slice sampler assuming uniform priors. Note that different parameters are present depending on the analysis. We list the lower and upper prior limit, the step size for sampling and the initial parameter value. For a subset of analyses, we run ten additional chains for convergence tests (see table~\ref{tab:sampling-gelman-rubin}) with random initial starting points drawn from the ``initial range'' in the last column.}
\label{tab:sampling-priors}
\end{table}

We select a subset of scenarios for a detailed convergence analysis, and for each we run ten additional chains whose starting points are drawn randomly from the flat interval given in table~\ref{tab:sampling-priors}. For each of these chains, we compute the Gelman-Rubin criterion which is listed in table~\ref{tab:sampling-gelman-rubin}. Overall, we find $\hat{R}-1 < 0.01$ with a very good margin for all scenarios explored, and hence no signs for poorly converged chains.

\newcolumntype{L}[1]{>{\raggedright\let\newline\\\arraybackslash\hspace{0pt}}m{#1}}
\begin{table}
\centering
\begin{tabular}{L{5.48in-7.cm}||p{1.2cm}|p{1.2cm}|p{1.2cm}|p{1.2cm}|p{1.2cm}}
\multirow[b]{2}{*}{data set and analysis} &
\multicolumn{4}{c}{Gelman-Rubin criterion $\hat{R} - 1$} \\
& $f$ & $\beta_{\partial_\pp\sigma}$ & $\sigma_0$ & $\sigma_{\epsilon,2}$ & $\sigma_{\epsilon\mu,2}$ \\
		\hline \hline
		2LPT, $o_\bias=3$ , $o_{u,\bias}=1$ (figure~\ref{fig:inference__onestep-varLPT-allz}), \newline
		M1, sim. 1, $z=0.0$, $\Lambda = 0.10\,h/\mpc$
		& 0.0004 & 0.0007 & 0.0042 & 0.0057 & 0.0008 \\
		\hline
		2LPT, $o_\bias=3$ , $o_{u,\bias}=1$ (figure~\ref{fig:inference__onestep-varLPT-allz}), \newline
		M2, sim. 1, $z=0.0$, $\Lambda = 0.10\,h/\mpc$
		& 0.0001 & $< 10^{-4}$ & 0.0022 &  0.0027 & 0.0002 \\
		\hline
		2LPT, $o_\bias=3$ , $o_{u,\bias}=1$ (figure~\ref{fig:inference__onestep-varLPT-allz}), \newline
		M1, sim. 1, $z=0.0$, $\Lambda = 0.18\,h/\mpc$
		& 0.0002 & 0.0002 & 0.0040 & 0.0035 & 0.0002 \\
		\hline
		2LPT, $o_\bias=3$ , $o_{u,\bias}=1$ (figure~\ref{fig:inference__onestep-varLPT-allz}), \newline
		M2, sim. 1, $z=0.0$, $\Lambda = 0.18\,h/\mpc$
		& $< 10^{-4}$ & $<10^{-4}$ & 0.0015 & 0.0018 & 0.007 \\
		\hline
		2LPT, $o_\bias=3$ , $o_{u,\bias}=1$ (figure~\ref{fig:inference__onestep-varLPT-allz}), \newline
		M1, sim. 1, $z=1.0$, $\Lambda = 0.10\,h/\mpc$
		& 0.0006 & 0.0007 & 0.0059 & 0.0058 & 0.0005\\
		\hline
		2LPT, $o_\bias=3$ , $o_{u,\bias}=1$ (figure~\ref{fig:inference__onestep-varLPT-allz}), \newline
		M2, sim. 1, $z=1.0$, $\Lambda = 0.10\,h/\mpc$
		& 0.0004 & 0.0007 & 0.0014 & 0.0020 & 0.0005 \\
		\hline
		2LPT, $o_\bias=3$ , $o_{u,\bias}=1$ (figure~\ref{fig:inference__onestep-varLPT-allz}), \newline
		M1, sim. 1, $z=1.0$, $\Lambda = 0.18\,h/\mpc$
		& $<10^{-4}$ & $<10^{-4}$ & 0.0041 & 0.0031 & 0.0003 \\
		\hline
		2LPT, $o_\bias=3$ , $o_{u,\bias}=1$ (figure~\ref{fig:inference__onestep-varLPT-allz}), \newline
		M2, sim. 1, $z=1.0$, $\Lambda = 0.18\,h/\mpc$
		& 0.0001 & 0.0002 & 0.0047 & 0.0051 & 0.0004 \\
		\hline
		3LPT, $o_\bias=3$ , $o_{u,\bias}=1$ (figure~\ref{fig:inference__onestep-varLPT-allz}), \newline
		M1, sim. 1, $z=0.0$, $\Lambda = 0.10\,h/\mpc$
		& 0.0005 & 0.0003 & 0.0014 & 0.0019 & 0.0009 \\
		\hline
		3LPT, $o_\bias=3$ , $o_{u,\bias}=1$ (figure~\ref{fig:inference__onestep-varLPT-allz}), \newline
		M2, sim. 1, $z=0.0$, $\Lambda = 0.10\,h/\mpc$
		& 0.0001 & $<10^{-4}$ & 0.0010 & 0.0011 & $<10^{-4}$\\
		\hline
		3LPT, $o_\bias=3$ , $o_{u,\bias}=1$ (figure~\ref{fig:inference__onestep-varLPT-allz}), \newline
		M1, sim. 1, $z=0.0$, $\Lambda = 0.18\,h/\mpc$
		& $<10^{-4}$ & 0.0001 & 0.0013 & 0.0015 & $<10^{-4}$\\
		\hline
		3LPT, $o_\bias=3$ , $o_{u,\bias}=1$ (figure~\ref{fig:inference__onestep-varLPT-allz}), \newline
		M2, sim. 1, $z=0.0$, $\Lambda = 0.18\,h/\mpc$
		& $<10^{-4}$ & $<10^{-4}$ & 0.0059 & 0.0053 & 0.0002 \\
		\hline
		3LPT, $o_\bias=3$ , $o_{u,\bias}=1$ (figure~\ref{fig:inference__onestep-varLPT-allz}), \newline
		M1, sim. 1, $z=1.0$, $\Lambda = 0.10\,h/\mpc$
		& 0.0007 & 0.0004 & 0.0013 & 0.0012 & 0.0003 \\
		\hline
		3LPT, $o_\bias=3$ , $o_{u,\bias}=1$ (figure~\ref{fig:inference__onestep-varLPT-allz}), \newline
		M2, sim. 1, $z=1.0$, $\Lambda = 0.10\,h/\mpc$
		& $<10^{-4}$ & $<10^{-4}$ & 0.0019 & 0.0025 & 0.0001 \\
		\hline
		3LPT, $o_\bias=3$ , $o_{u,\bias}=1$ (figure~\ref{fig:inference__onestep-varLPT-allz}), \newline
		M1, sim. 1, $z=1.0$, $\Lambda = 0.18\,h/\mpc$
		& $<10^{-4}$ & $<10^{-4}$ & 0.0029 & 0.0027 & 0.0001 \\
		\hline
		3LPT, $o_\bias=3$ , $o_{u,\bias}=1$ (figure~\ref{fig:inference__onestep-varLPT-allz}), \newline
		M2, sim. 1, $z=1.0$, $\Lambda = 0.18\,h/\mpc$
		& $<10^{-4}$ & $<10^{-4}$ & 0.0019 & 0.0029 & 0.0010 \\
		\hline
		3LPT, $o_\bias=3$ w/ $\partial_\pp^2\sigma$ (figure~\ref{fig:inference__onestep-velocitybias}), \newline
		M1, sim. 1, $z=0.0$, $\Lambda = 0.10\,h/\mpc$
		& $<10^{-4}$ & -- & 0.0021 & 0.0028 & 0.0011 \\
		\hline
		3LPT, $o_\bias=3$ w/ $\partial_\pp^2\sigma$ (figure~\ref{fig:inference__onestep-velocitybias}), \newline
		M2, sim. 1, $z=0.0$, $\Lambda = 0.10\,h/\mpc$
		& 0.0001 & -- & 0.0028 & 0.0033 & 0.0002 \\
		\hline
		3LPT, $o_\bias=3$ w/ $\partial_\pp^2\sigma$ (figure~\ref{fig:inference__onestep-velocitybias}), \newline
		M1, sim. 1, $z=0.0$, $\Lambda = 0.18\,h/\mpc$
		& $<10^{-4}$ & -- & 0.0045 & 0.0043 & 0.0005 \\
		\hline
		3LPT, $o_\bias=3$ w/ $\partial_\pp^2\sigma$ (figure~\ref{fig:inference__onestep-velocitybias}), \newline
		M2, sim. 1, $z=0.0$, $\Lambda = 0.18\,h/\mpc$
		& $<10^{-4}$ & -- & 0.0037 & 0.0054 & 0.0013 \\
	\end{tabular}
	\caption{Convergence test for selected analyses. The lower and higher mass bins are abbreviated as ``M1'' and ``M2'', respectively. We list the Gelman-Rubin statistics from 10 chains with 2,000 samples each after discarding 100 samples for burn in. Note that $b_{\partial_\pp^2\sigma}$ in the last four scenarios is marginalized analytically and hence does not appear explicitly in the list.}
	\label{tab:sampling-gelman-rubin}
\end{table}

\begin{figure}
\centering
\includegraphics[]{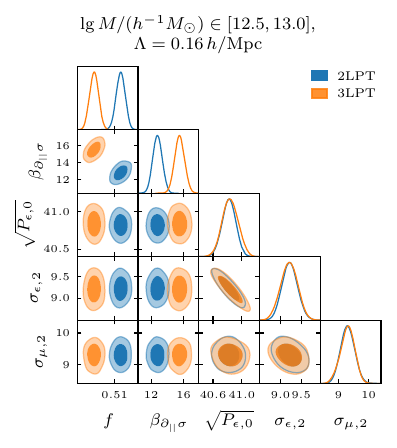}%
\includegraphics[]{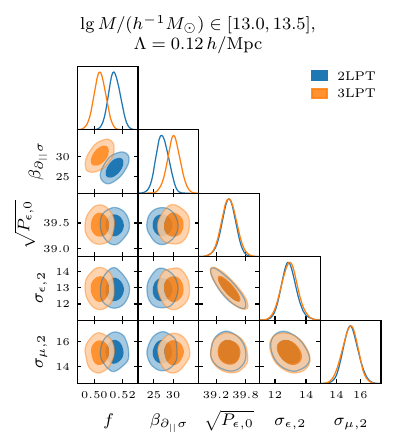}
\caption{Contour plots of all parameters which are varied in an analysis, here for the one-step model at $z=0.0$. For the lower mass bin (\textbf{left}) we show results at $\Lambda=0.12\,h/\mpc$ and for the higher one (\textbf{right}) at $\Lambda=0.16\,h/\mpc$, each time comparing the 2LPT and the 3LPT forward model with third order bias and the leading velocity operator. The corresponding one-dimensional constraints on $f$ are shown in figure~\ref{fig:inference__onestep-varLPT-allz}. }
\label{fig:sampling_contours}
\end{figure}

We report the final parameter constraints of our analysis in terms of the mode-centered 68\% credible interval. These agree well with the mean-centered $1\sigma$ limits and also the 95\% credible intervals overlap well with the $2\sigma$ constraint on $f$, indicating that its posterior is close to Gaussian. In figure~\ref{fig:sampling_contours} we provide a more detailed view on the posteriors of some selected results from our baseline analysis in figure~\ref{fig:inference__onestep-varLPT-allz}, showing the two-dimensional contours\footnote{Contour plots are generated with \texttt{GetDist}, \url{https://getdist.readthedocs.io/}} of all free parameters.

\section{Supplementary tests}
\label{sec: additional-tests}

\subsection{Accuracy of the forward model}
\label{sec: additional-tests-modelaccuracy}

\begin{figure}
\centering
\includegraphics[]{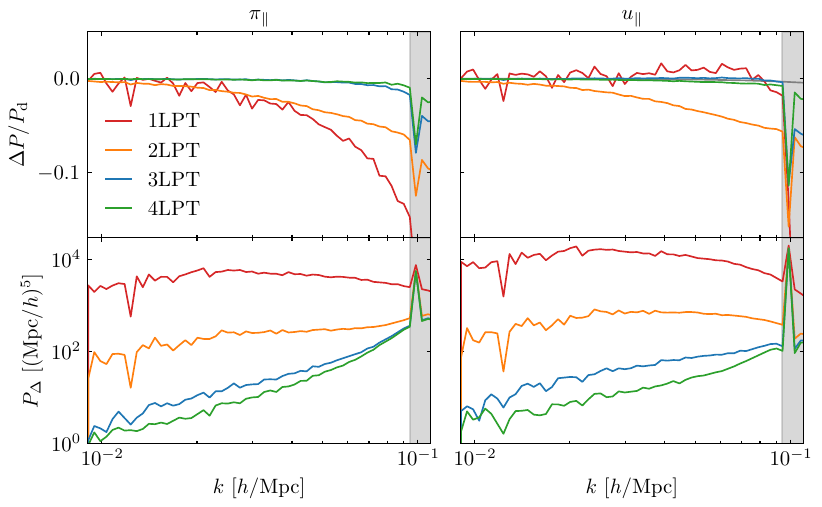}
\caption{Perturbative accuracy of the forward model for the momentum density $\pi_\pp$ (\textbf{left}) and the velocity $u_\pp$ (\textbf{right}) along the LOS-direction at $z=0.5$. We compare results from the forward model to a reference N-body simulation which has an identical Fourier-space top-hat filter at $\Lambda=0.10\,h/\mpc$ applied to its initial conditions. In the top row we show the residuals of the power spectra and in the bottom row the residual power spectra. We compute the velocity on a grid of size $N_\mathrm{G}=256$ from simulations with $N_\mathrm{part.}=1536$ and the forward model with $N_\mathrm{part.}=1024$ particles using the Delaunay tessellation algorithm. Gray lines indicate the expected difference due to the the different number of particles in forward model and simulations. The 4LPT mode predicts the power spectrum to better than 1\% accuracy for momentum and velocity. This extends the results of figures~\ref{fig:velocity-accuracy__perturbative-accuracy-momentum} and~\ref{fig:velocity-accuracy__perturbative-accuracy-velocity} to lower cut-offs.}
\label{fig:velocity-accuracy__accuracy-velocity-momentum-density_L010_z050}
\end{figure}

\begin{figure}
\centering
\includegraphics[]{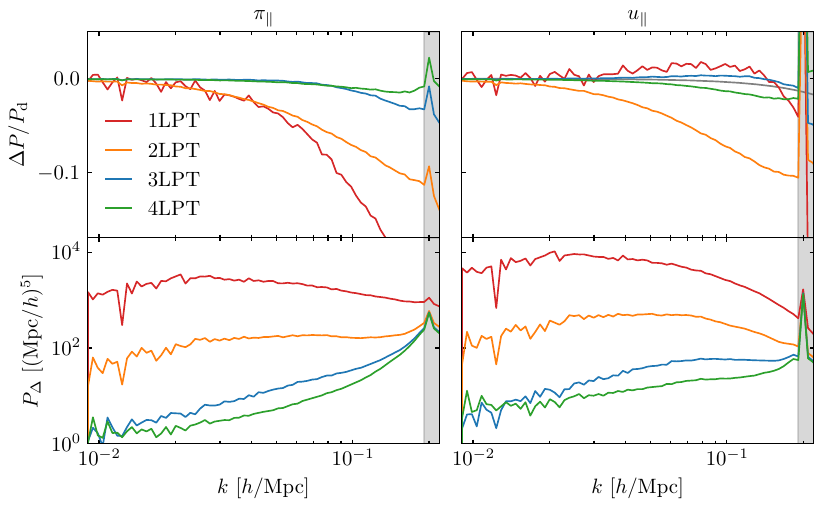}
\caption{Perturbative accuracy of the forward model for the momentum density $\pi_\pp$ (\textbf{left}) and the velocity $u_\pp$ (\textbf{right}) along the LOS-direction at $z=0.5$. Similar to figure~\ref{fig:velocity-accuracy__accuracy-velocity-momentum-density_L010_z050} but now for $z=1.0$ and $\Lambda=0.20\,h/\mpc$. The 4LPT mode predicts the power spectrum to better than 1\% accuracy for momentum and velocity. This extends the results of figures~\ref{fig:velocity-accuracy__perturbative-accuracy-momentum} and~\ref{fig:velocity-accuracy__perturbative-accuracy-velocity} to higher redshifts.}
\label{fig:velocity-accuracy__accuracy-velocity-momentum-density_z100}
\end{figure}

\begin{figure}
\centering
\includegraphics[]{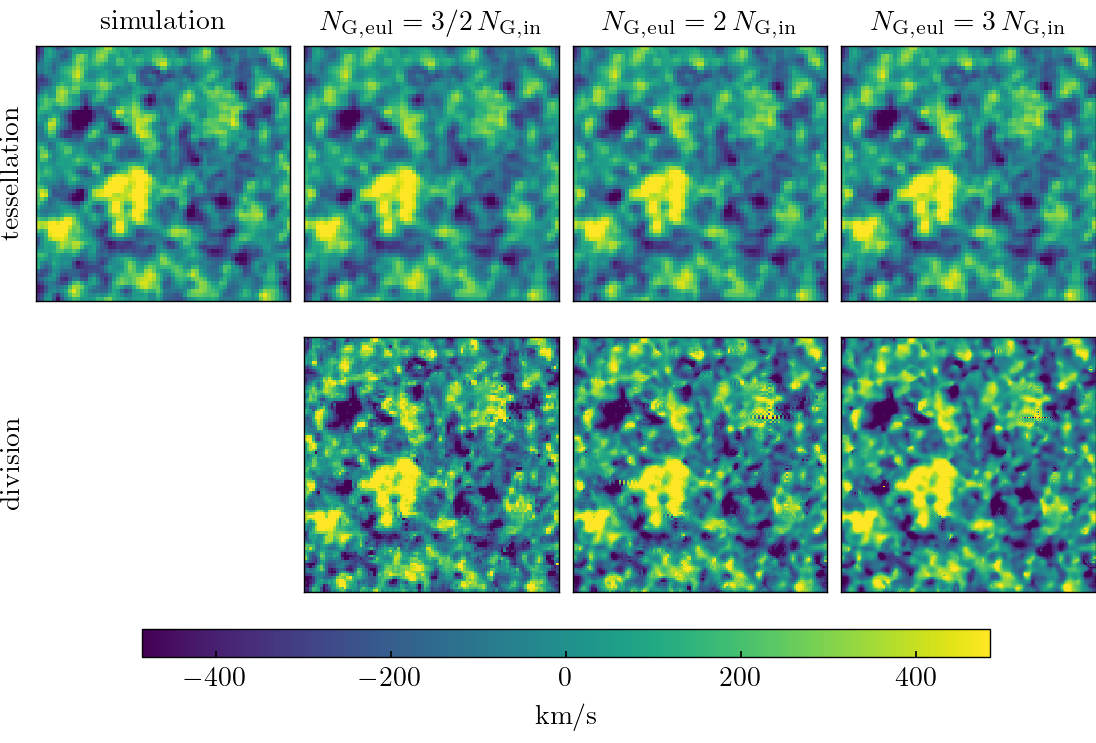}
\caption{Slices through the LOS-velocity $v_\pp$ at $z=0.0$ for a cut-off $\Lambda=0.10\,h/\mpc$ where the LOS direction is perpendicular to the image plane. In the top row we show velocity grids computed by the Delaunay tessellation for a grid size of $N_\delaunay=64$ from the reference simulation (left) and from the forward model at increasing particle numbers $N_\eul$. We compare these to velocity grids found by dividing out the density from the momentum in the bottom panels.}
\label{fig:velocity-accuracy__presentation__grid-comparison-L010}
\end{figure}

\begin{figure}
\centering
\includegraphics[]{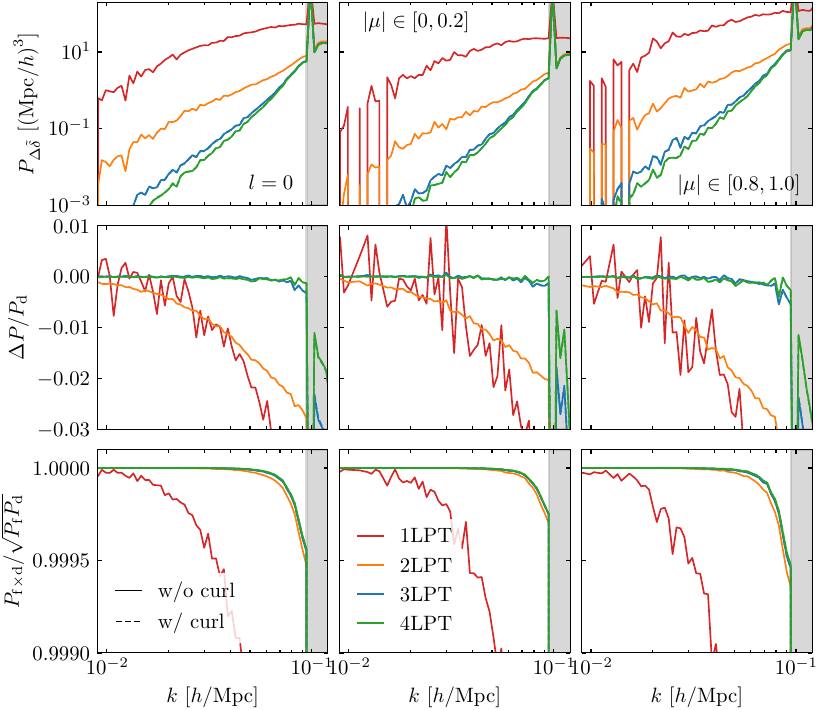}
\caption{The perturbative error of the redshift-space forward model at $z=0.5$. We compare the one-step forward model at different LPT (subscript ``m'') orders to a reference N-body simulation (subscript ``s'') started from identical initial conditions with a cut-off at $\Lambda=0.10\,h/\mpc$. The comparison is shown in terms of the residual power spectrum (top row), the power spectrum difference (middle row) and the cross-correlation between model and reference simulation (bottom row) for the power spectrum monopole (left) and two power spectrum wedges perpendicular (middle) and parallel (right) to the line of sight. For LPT orders $n_\lpt \geq 3$, we distinguish between results with and without the transverse displacement contribution, however, the impact is negligible. This extends the results of figure~\ref{fig:redshiftspace-accuracy__lpterror_z050} to lower cut-offs.}
\label{fig:redshiftspace-accuracy__lpterror_L010_z050}
\end{figure}

We present additional tests on the numerical and the physical convergence of the forward model, in particular considering the following scenarios.
\begin{itemize}
\item The perturbative accuracy of momentum and velocity is explored in figures~\ref{fig:velocity-accuracy__perturbative-accuracy-momentum} and~\ref{fig:velocity-accuracy__perturbative-accuracy-velocity}, respectively, for $\Lambda=0.20\,h/\mpc$ and $z=0.5$. We extend this test to a lower cut-off, $\Lambda=0.10\,h/\mpc$, and higher redshifts $z=1.0$ in figures~\ref{fig:velocity-accuracy__accuracy-velocity-momentum-density_L010_z050} and~\ref{fig:velocity-accuracy__accuracy-velocity-momentum-density_z100}.
\item One slice through the LOS velocity is compared between reference simulation and forward model in figure~\ref{fig:velocity-accuracy__presentation__grid-comparison-L020} for a cut-off $\Lambda=0.20\,h/\mpc$. In particular, this comparison considers different ways of computing the volume weighted velocity assignment and explores different numerical resolutions. In figure~\ref{fig:velocity-accuracy__presentation__grid-comparison-L010} we extend this analysis to a lower cut-off, $\Lambda=0.10\,h/\mpc$.
\item The analysis of the perturbative accuracy of the redshift-space forward model, shown in figure~\ref{fig:redshiftspace-accuracy__lpterror_z050} for $\Lambda=0.20\,h/\mpc$ and $z=0.5$, is extended to a lower cut-off, $\Lambda=0.10\,h/\mpc$ in figure~\ref{fig:redshiftspace-accuracy__lpterror_L010_z050}.
\end{itemize}

\subsection{Parameter inference at fixed initial conditions}
\label{sec:additional-tests-inference}
\begin{figure}
\centering
\includegraphics[]{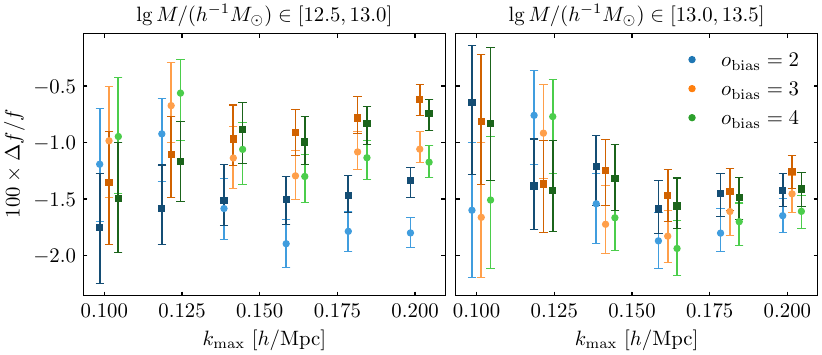}
\caption{Inference of the growth rate $f$ from N-body halos at $z=0$ assuming fixed initial conditions with the 3LPT one-step model for gravity, third order Lagrangian bias and $\Lambda=k_\mathrm{max}$. There are two realizations of the simulations available, which are analyzed independently and indicated by different symbols. We explore the impact of different orders in the bias expansion, always considering the leading-order velocity bias term. In contrast to the rest-frame analysis of $\sigma_8$ \cite{Stadler:2024}, the bias order only has a minor impact on the inferred values of $f$.}
\label{fig:inference__onestep-3lpt-varbiasorder}
\end{figure}

\begin{figure}
\centering
\includegraphics[]{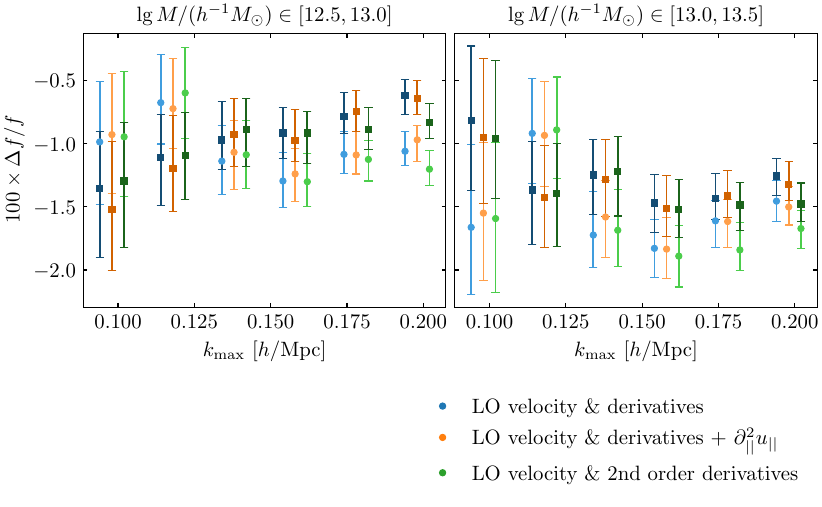}
\caption{Inference of the growth rate $f$ from N-body halos at $z=0$ assuming fixed initial conditions with the 3LPT one-step model for gravity, third order Lagrangian bias and $\Lambda=k_\mathrm{max}$. Two realizations of the simulations are analyzed independently and indicated by different symbols. We compare the baseline results from figure~\ref{fig:inference__onestep-varLPT-allz} to two model extensions with higher-derivative operators. The $\partial_\pp^2 u_\pp$ term in the velocities enters the redshift-space density contrast as $\mu^4 k^2 \delta$ at leading order. Further, we explore higher derivative operators in the density expansion that are constructed from applying the Laplace operator to the second-order bias terms. Neither of these extensions has a significant impact on the analysis.}
\label{fig:inference__onestep-higherscaleseffects}
\end{figure}

\begin{figure}
\centering
\includegraphics[]{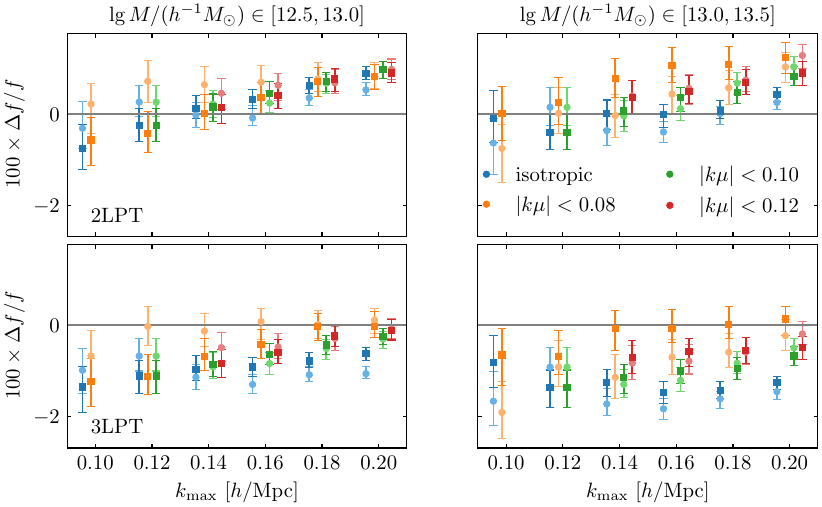}
\caption{Inference of the growth rate $f$ from N-body halos at $z=0$ assuming fixed initial conditions with the 3LPT one-step model for gravity and third order Lagrangian bias. Two realizations of the simulations are analyzed independently and indicated by different symbols. Perpendicular to the LOS, we generally choose $\Lambda = k_\mathrm{max}$, and we test more aggressive filters along the LOS direction. The filter scale in the legend is quoted in units of $h/\mpc$. In particular for the 3LPT model, removing high-k modes close to the LOS (i.e. $\mu$ close to one) can improve the agreement between inferred growth rate and ground truth.}
\label{fig:inference__onestep-anisotropicfilter}
\end{figure}

\begin{figure}
\centering
\includegraphics[]{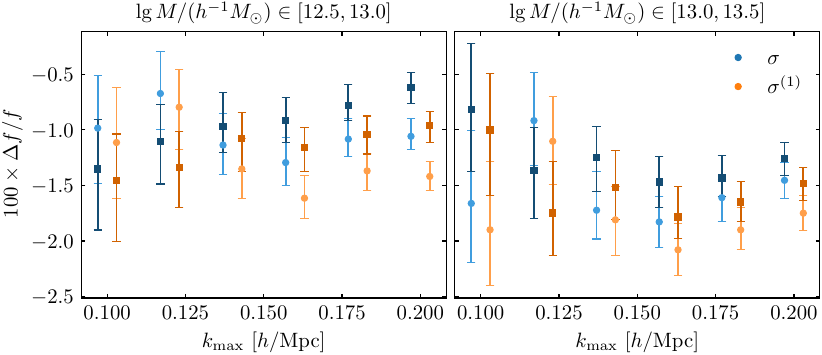}
\caption{Inference of the growth rate $f$ from N-body halos at $z=0$ assuming fixed initial conditions with the 3LPT one-step model for gravity, third order Lagrangian bias and $\Lambda=k_\mathrm{max}$. Two realizations of the simulations are analyzed independently and indicated by different symbols. We compare results where the bias expansion is constructed from $\sigma=\sum_n \sigma^{(n)}$ to using $\sigma^{(1)}$ instead. Up to terms beyond the order of the bias expansion, $o_\mathrm{bias}$, both options should be equivalent and indeed we find no significant difference.}
\label{fig:inference__sigma-crosschekc}
\end{figure}

\begin{figure}
\centering
\includegraphics[]{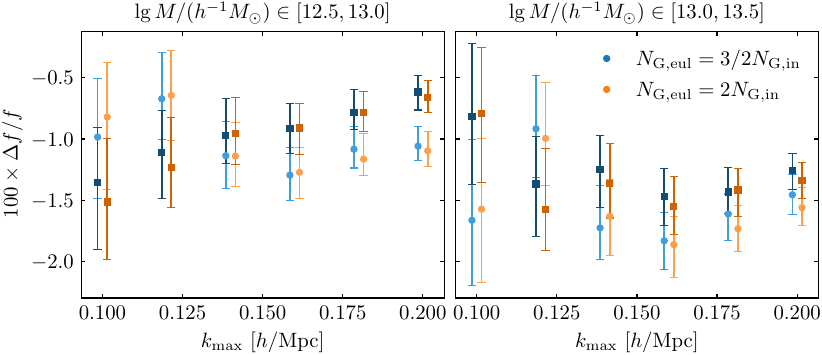}
\caption{Inference of the growth rate $f$ from N-body halos at $z=0$ assuming fixed initial conditions using the one-step model with 3LPT gravity and $\Lambda=k_\mathrm{max}$. There are two realizations of the simulations available, which are analyzed independently and indicated by different symbols. We test if residual resolution effects might still impact the analysis, by increasing $N_\eul$ to $2 N_\ini$, and compare these results to the baseline 3LPT results from figure~\ref{fig:inference__onestep-varLPT-allz}. We find no evidence that a too low numerical resolution should impact the baseline result.}
\label{fig:inference__onestep-ngeul}
\end{figure}

\begin{figure}
\centering
\includegraphics[]{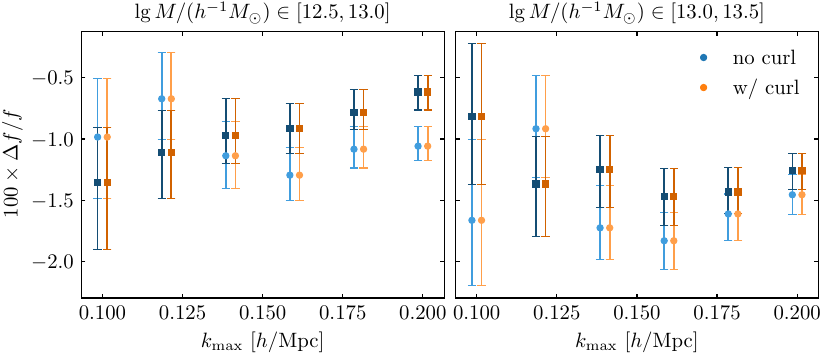}
\caption{Inference of the growth rate $f$ from N-body halos at $z=0$ assuming fixed initial conditions with the 3LPT one-step model for gravity, third order Lagrangian bias and $\Lambda=k_\mathrm{max}$. Two realizations of the simulations are analyzed independently and indicated by different symbols. We compare results where the transverse contribution to the LPT displacement vector is neglected (``no curl'') to those where it is included, and we find that it has no impact on the analysis.}
\label{fig:inference__onestep-curl}
\end{figure}

We show additional parameter recovery tests in figures~\ref{fig:inference__onestep-3lpt-varbiasorder} to~\ref{fig:inference__onestep-curl} and comment on them in section~\ref{sec:results_results}.

\clearpage
\bibliographystyle{JHEP} 
\bibliography{rsd-speed-and-accuracy.bib}

\end{document}